\documentclass{article}

\usepackage{arxiv}

\usepackage[utf8]{inputenc} % allow utf-8 input
\usepackage[T1]{fontenc}    % use 8-bit T1 fonts
\usepackage{hyperref}       % hyperlinks
\usepackage{url}            % simple URL typesetting
\usepackage{booktabs}       % professional-quality tables
\usepackage{amsfonts}       % blackboard math symbols
\usepackage{nicefrac}       % compact symbols for 1/2, etc.
\usepackage{microtype}      % microtypography
\usepackage{lipsum}		% Can be removed after putting your text content
\usepackage{graphicx}
\usepackage{natbib}        % style references
\usepackage{doi}
\usepackage{mhchem}        % chemistry formula

\title{Capability demonstration of a JEDI-based system for TEMPO assimilation: system description and evaluation}

\author{
\href{https://orcid.org/0000-0002-5223-9238}{\includegraphics[scale=0.06]{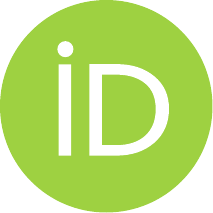}\hspace{1mm}Maryam Abdi-Oskouei}\thanks{Corresponding author: Maryam Abdi-Oskouei, \texttt{maryam.abdi.oskouei@gmail.com}} \\
JCSDA/UCAR
\and
\href{https://orcid.org/0009-0007-5691-5540}{\includegraphics[scale=0.06]{orcid.pdf}\hspace{1mm}Jérôme Barré} \\
JCSDA/UCAR
}

\date{}

\renewcommand{\shorttitle}{JEDI-Based assimilation for TEMPO}

\hypersetup{
pdftitle={JEDI-Based DA system for TEMPO assimilation},
pdfsubject={Atmospheric composition data assimilation},
pdfauthor={Maryam Abdi-Oskouei and Jérôme Barré},
pdfkeywords={data assimilation, atmospheric composition, TEMPO},
}

\begin{document}
\maketitle

\begin{abstract}

The launch of the Tropospheric Emissions: Monitoring of Pollution (TEMPO) mission in 2023 marked a new era in air quality monitoring by providing high-frequency, geostationary observations of column $\mathrm{NO}_2$ across North America. In this study, we present the first implementation of a TEMPO $\mathrm{NO}_2$ data assimilation system using the Joint Effort for Data assimilation Integration (JEDI) framework. Leveraging a four-dimensional ensemble variational (4DEnVar) approach and an Ensemble of Data Assimilations (EDA), we demonstrate a novel capability to assimilate hourly $\mathrm{NO}_2$ retrievals from TEMPO alongside polar-orbiting TROPOMI data into NASA’s GEOS Composition Forecast (GEOS-CF) model. The system is evaluated over the CONUS region for August 2023, using a suite of independent measurements including Pandora spectrometers, AirNow surface stations, and aircraft-based observations from AEROMMA and STAQS field campaigns. Results show that the assimilation system successfully integrates geostationary $\mathrm{NO}_2$ observations, improves model performance in the column, and captures diurnal variability. However, assimilation also leads to systematic reductions in surface $\mathrm{NO}_2$ levels, improving agreement with some datasets (e.g., Pandora, AEROMMA) but degrading comparisons with others (e.g., AirNow). These findings highlight the importance of joint evaluation across platforms and motivate further development of dual-concentration emission assimilation schemes. While the system imposes high computational costs, primarily from the forecast model, ongoing efforts to integrate AI-based model emulators offer a promising path toward scalable, real-time assimilation of geostationary atmospheric composition data.
\end{abstract}

\section{Introduction}
\label{sec:introduction}

The advent of geostationary (GEO) satellite observations is revolutionizing our ability to monitor air quality (AQ) and atmospheric composition with high temporal resolution. Unlike polar-orbiting satellites, which provide global coverage but only observe a given location once or twice daily, geostationary satellites offer continuous regional monitoring. This capability supports hourly or sub-hourly measurements critical for understanding diurnal variations in emissions and chemical transformations.

International efforts have led to the development of a constellation of GEO satellites dedicated to air quality monitoring, coordinated by the Committee on Earth Observation Satellites (CEOS) Atmospheric Composition Virtual Constellation as early as 2006 \citep{ceos_atmospheric_composition_constellation_geostationary_2011} and then more formulated in CEOS, 2019 \citep{ceos_atmospheric_compositin_virtual_constellation_geostationary_2019}. This GEO constellation includes the Geostationary Environment Monitoring Spectrometer (GEMS) over Asia, the Tropospheric Emissions: Monitoring of Pollution (TEMPO) mission over North America, and the Sentinel-4 instrument over Europe. These missions aim to provide complementary, continuous air quality measurements, advancing our ability to monitor and predict atmospheric pollution on a regional and global scale \citep{barre_feasibility_2015, barre_feasibility_2016}.
The Geostationary Coastal and Air Pollution Events (GEO-CAPE) initiative, a mission concept developed by NASA to improve observations of coastal ocean dynamics and air pollution over North America, has recognized the importance of the geostationary atmospheric composition monitoring over North America \citep{fishman_united_2012, al-saadi_advancing_2018}. Although GEO-CAPE was never fully realized as a single mission, its scientific and technical foundations have influenced the subsequent geostationary air quality satellite missions mentioned above. GEMS, launched in 2020, has already demonstrated its capability to monitor diurnal variations in key pollutants such as nitrogen dioxide ($\mathrm{NO}_2$), ozone (\ce{O3}), and formaldehyde (\ce{HCHO}), providing insights into emission sources and atmospheric chemistry processes \citep{kim_new_2020, edwards_quantifying_2024}.

TEMPO, launched in 2023, is the first instrument on a geostationary orbit dedicated to air quality monitoring over North America. The instrument provides hourly measurements of key atmospheric pollutants, including $\mathrm{NO}_2$, ozone, and formaldehyde, with unprecedented spatial and temporal resolution \citep{zoogman_tropospheric_2017}. TEMPO’s ability to measure rapid changes in pollution levels due to traffic emissions, industrial activity, and meteorological factors marks a significant advancement in air quality monitoring and forecasting. Data assimilation (DA) is essential to integrate these observations into numerical models, improving predictions and providing more accurate and actionable air quality assessments.

Despite the promise of GEO air quality observations, existing atmospheric composition DA systems face significant challenges in assimilating high-frequency geostationary data. The Kalman filters, such as in \citet{hsu_observing_2024} and \citet{miyazaki_balance_2019, miyazaki_evaluation_2020}, use the flow-dependent information of the ensemble but lack resolving time-evolving information within the data assimilation window. \citet{hsu_observing_2024} showed that the usually used 6-hourly assimilation window should be reduced to a 3-hourly window but would still lack resolving the 1-hour resolution of geostationary instruments and, therefore, not make the best use of observations. Moreover, going toward finer windows to match the GEO temporal resolution can drastically slow down the throughput of such systems (i.e., increasing queuing on high-performance computers).  

The High-Resolution Rapid Refresh (HRRR) is an operational convection-allowing numerical weather prediction system that provides forecasts by assimilating a broad range of observations using 1-hourly assimilation windows. These hourly updates enable for more accurate short-range forecasts \citep{james_high-resolution_2022,benjamin_north_2016}. However, the computational demands of running at 3~km resolution every hour limit HRRR’s application to regional domains, such as the continental United States (CONUS). Its cost and throughput are not suitable for a fully comprehensive air quality system that integrates nitrogen oxides (\(\mathrm{NO}_x\)) and volatile organic compound (VOC) chemistry, especially when running on lower-priority queues on operational high-performance computing (HPC) systems.

In contrast, four-dimensional variational (4D-Var) assimilation approaches can better capture time-evolving atmospheric states without the need to reduce the assimilation window length; such a method allows for resolution of the time-varying information within the same data assimilation window but requires coding and maintenance of a tangent linear and adjoint of the forecast model, making this methodology less flexible compared to other assimilation methods (e.g., Copernicus Atmosphere Monitoring Service (CAMS) \citep{inness_cams_2019} \ and the GEOS-Chem adjoint system \citep{qu_sectorbased_2022}).

The Joint Effort for Data assimilation Integration (JEDI) framework \citep{tremolet_joint_2020} provides a generic and flexible approach for assimilating GEO observations. Among multiple data assimilation “flavors”, such as 3DVar and 3DEnVar \citep{liu_data_2022, guerrette_data_2023, huang_jedi-based_2023, wei_prototype_2024}, JEDI supports a four-dimensional ensemble variational (4DEnVar) assimilation framework that is model-agnostic and does not require tangent-linear or adjoint models. The 4DEnVar methodology leverages ensemble information to capture time dependence within the assimilation window \citep{bowler_effect_2017}, making it well-suited for high-temporal-resolution data assimilation from GEO instruments targeting atmospheric composition.

The goal of this work is to demonstrate a novel GEO data assimilation capability for atmospheric composition using the JEDI framework. We employ NASA’s Goddard Earth Observing System Composition Forecast (GEOS-CF) model \citep{keller_description_2021} as a testbed to showcase the feasibility and benefits of assimilating high-frequency GEO data. While GEOS-CF is used in this study, the JEDI system is designed to be generic and adaptable to other models, highlighting its potential for broader applications in atmospheric composition data assimilation. By leveraging JEDI’s advanced 4DEnVar capabilities, this work aims to advance the integration of TEMPO and other GEO satellite data into air quality forecasting and emission estimation systems, ultimately improving our ability to monitor and predict air pollution at unprecedented temporal and spatial scales.

As with any new satellite observing system, evaluating the quality and utility of TEMPO retrievals is a critical step toward their effective integration into atmospheric composition models and forecasting systems. Geostationary instruments such as TEMPO introduce novel observational characteristics, including high temporal resolution, variable viewing geometry, and potentially complex retrieval biases, which require systematic assessment. Traditionally, studies that compare satellite products to ground-based or airborne observations rely on spatial and temporal co-location strategies to account for differences in measurement footprint, vertical sensitivity, and sampling frequency. However, these co-location approaches are often limited by the heterogeneity of the datasets and may not fully resolve underlying discrepancies. In contrast, data assimilation offers a physically consistent framework to evaluate observational quality, particularly when using flow-dependent background error covariances that propagate observational information in space and time. By integrating observations within a dynamically constrained modeling environment, assimilation can reveal systematic biases, enhance observational representativeness, and facilitate robust intercomparison across platforms. This makes data assimilation a uniquely powerful tool for validating and extracting value from new satellite datasets such as TEMPO.

This paper focuses on the description and demonstration of the JEDI system’s capability to assimilate high-frequency satellite observations, specifically TEMPO and TROPOMI $\mathrm{NO}_2$ retrievals, into a state-of-the-art chemical forecasting system. Beyond presenting the assimilation system itself, we also emphasize the breadth of evaluation and validation that JEDI enables through its unified framework, including comparisons to multiple independent datasets across platforms and observing geometries. To our knowledge, this is the first study to implement and demonstrate an ensemble of data assimilation (EDA;\citep{isaksen_ensemble_2010}) in combination with four-dimensional ensemble variational (4DEnVar) methods for air quality applications. This novel configuration allows for both flow-dependent background error representation and robust uncertainty quantification, marking a significant advancement in chemical data assimilation methodology.

The paper is structured as follows: Section~\ref{sec:methods} describes the assimilated and independent datasets, as well as the configuration of the JEDI system, including the observation operators, background error representation, assimilation algorithm, and forecast model setup. Section~\ref{sec:results} presents results for a set of diagnostics in both the observation and the model space, with a focus on diurnal variability, increment structure, and ensemble behavior. In addition, a thorough validation against independent datasets is provided. Section~\ref{sec:discussion} offers a broader discussion and conclusion on system performance, limitations, and key insights into observation–model interactions, motivating further developments and studies.

\section{Methods}
\label{sec:methods}

\subsection{Assimilated Observations}
\label{subsec:assimilated_observations}

\begin{figure}
    \centering
    \includegraphics[width=0.75\linewidth]{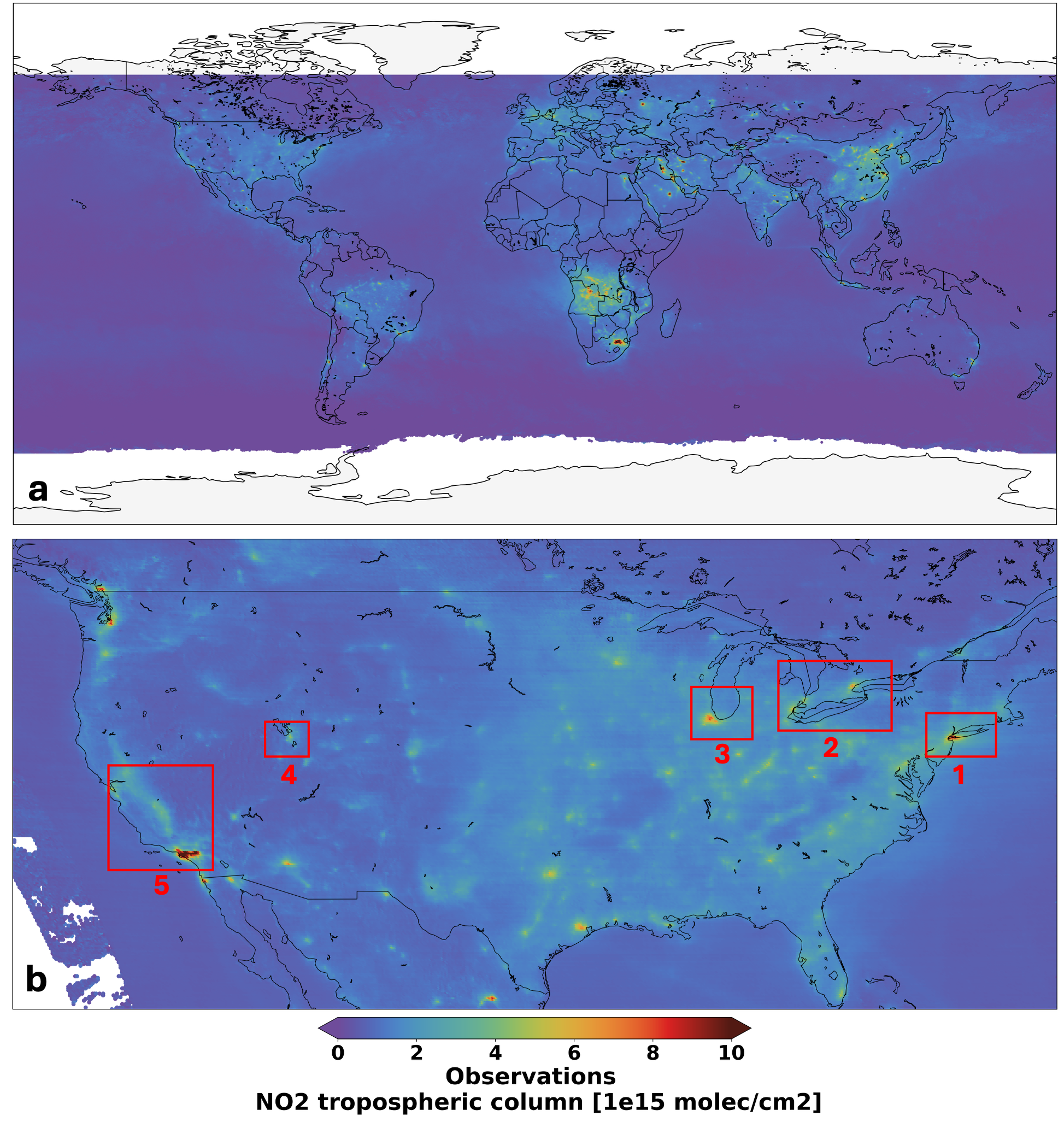}
    \caption{Mean tropospheric \ce{NO2} column during 4 to 31 August 2023, measured by a) TROPOMI instrument, and b) TEMPO. Red boxes 1-Long Island, 2-Toronto, 3-Lake Michigan, 4-Salt Lake City, and 5-California indicate the study regions defined to encompass flight tracks from AEROMMA and STAQS field campaigns.}
    \label{fig:mean_obs_no2}
\end{figure}

Tropospheric $\mathrm{NO}_2$ column observations from the TROPOspheric Monitoring Instrument (TROPOMI) and the Tropospheric Emissions: Monitoring of Pollution (TEMPO) instrument were assimilated in this study. Figure~\ref{fig:mean_obs_no2} shows the mean tropospheric column $\mathrm{NO}_2$ assimilated into the data assimilation experiment from TROPOMI~(a) and TEMPO~(b) during the experiment period (4–31 August 2023). The observational data points were regridded to \(0.2^\circ \times 0.2^\circ\) for TROPOMI and \(0.1^\circ \times 0.1^\circ\) for TEMPO before averaging. While TROPOMI provides once-daily global coverage, TEMPO offers high-frequency measurements throughout the daytime over North America. Thus, the main analysis in this study focuses on the CONUS domain.

Red boxes in Figure~\ref{fig:mean_obs_no2}b indicate study regions that correspond to the field campaign flight tracks (details in Section~\ref{subsubsec:aircraft}), which are used as independent datasets to evaluate the impact of the assimilation. These study regions are used throughout the paper to assess assimilation impacts at the regional scale. Most are located in populated areas with significant urban emissions. The Salt Lake City study region (Box4 in Fig.\ref{fig:mean_obs_no2}b) covers a smaller area, while the California study region (Box5 in Fig.\ref{fig:mean_obs_no2}b) encompasses both urban and rural environments.

%\subsubsection{TEMPO tropospheric NO\textsubscript{2} column retrievals}
\subsubsection[**TEMPO tropospheric NO2 column retrievals**]{TEMPO tropospheric \ce{NO2} column retrievals}
\label{subsubsec:tempo_trop}

The TEMPO instrument is a geostationary ultraviolet-visible (UV–Vis) spectrometer developed to provide high-resolution, hourly measurements of atmospheric pollutants over North America at a spatial resolution of \(2.1 \times 4.4~\mathrm{km}^2\) \citep{zoogman_tropospheric_2017}. TEMPO operates in the 290–740~nm spectral range, enabling the detection of key trace gases such as $\mathrm{NO}_2$, ozone, and formaldehyde. Operating from a geostationary orbit at approximately \(91^\circ\)W longitude, TEMPO offers continuous daytime coverage from the Atlantic to the Pacific, and from approximately Mexico City to the Canadian oil sands. This configuration allows for unprecedented monitoring of diurnal variations in air quality across the continent.

The retrieval of $\mathrm{NO}_2$ concentrations from TEMPO measurements employs the Differential Optical Absorption Spectroscopy (DOAS) technique. This method involves analyzing the absorption features of $\mathrm{NO}_2$ in the measured spectra to determine slant column densities (SCDs). These SCDs are then converted to vertical column densities (VCDs) using air mass factors (AMFs) that account for factors such as viewing geometry, surface reflectance, and atmospheric conditions \citep{nowlan_nitrogen_2016}. The retrieval algorithm is designed to provide accurate and timely data to support air quality monitoring and research. For more details about the TEMPO retrieval algorithm, see the TEMPO Algorithm Theoretical Basis Document (ATBD) \citep{nowlan_tempo_2025}.

In this study, we used TEMPO tropospheric $\mathrm{NO}_2$ vertical column Level-2, Version 3 during August 2023. Following the recommendations in the TEMPO ATBD, we filtered out measurements with cloud fractions greater than 0.15 and solar zenith angles above \(70^\circ\).

%\subsubsection{TropOMI tropospheric NO\textsubscript{2} column retrievals}
\subsubsection[**TropOMI tropospheric NO2 column retrievals**]{TropOMI tropospheric \ce{NO2} column retrievals}
\label{subsubsec:tropomi}

TROPOMI is a nadir-viewing hyperspectral sensor onboard the Sentinel-5 Precursor satellite, launched in October 2017 as part of the Copernicus program. The instrument operates in the UV–Vis spectral range, with a spatial resolution of up to \(3.5 \times 5.5~\mathrm{km}^2\) (later improved to \(3.5 \times 3.5~\mathrm{km}^2\) in August 2019), enabling detailed mapping of $\mathrm{NO}_2$ pollution at the urban scale \citep{van_geffen_tropomi_2022}. The $\mathrm{NO}_2$ retrieval algorithm employs DOAS to derive slant column densities (SCDs), which are subsequently converted to vertical column densities (VCDs) using air mass factors (AMFs) derived from radiative transfer modeling and a priori profile information \citep{veefkind_tropomi_2012}. The product includes both tropospheric and stratospheric $\mathrm{NO}_2$ components, and it is routinely used for air quality monitoring, trend analysis, and validation of atmospheric chemistry and transport models. In the context of data assimilation, TROPOMI $\mathrm{NO}_2$ retrievals provide observational constraints on reactive nitrogen distributions and emissions, particularly in high-density source regions.

We used daily TROPOMI tropospheric $\mathrm{NO}_2$ vertical column Level-2 data from the offline dataset for August 2023. Only measurements with \texttt{qa\_value} greater than 0.75 were used, and data outside the latitude range of \(-65^\circ\) to \(65^\circ\) were excluded from assimilation.

\subsection{Independent Observations}
\label{subsec:indie_obs}

\subsubsection{AirNow}
\label{subsubsec:airnow}

The AirNow system, operated by the U.S. Environmental Protection Agency (EPA), provides near-real-time, surface-level air quality observations across the United States, focusing on key regulatory pollutants such as ozone and $\mathrm{NO}_2$. These observations are collected from a network of fixed monitoring stations operated by federal, state, tribal, and local air quality agencies. Data are reported hourly and undergo preliminary quality control procedures to ensure reliability for operational use and scientific analysis. In this study, AirNow ozone and $\mathrm{NO}_2$ observations are used as an independent benchmark for evaluating surface-level pollutant concentrations and assessing the performance of model simulations and data assimilation products.

To account for spatial variability and the representativeness of surface observations, monitoring sites are classified as urban, suburban, or rural based on EPA-provided metadata, including land use, population density, and proximity to emission sources. Urban sites are generally located in densely populated areas with high local emissions from traffic, industry, and residential activities, and are representative of high-concentration, source-influenced environments. Suburban sites are situated in intermediate zones, reflecting a mix of local and transported pollution, while rural sites are positioned in low-population areas with minimal local emissions and are more representative of background atmospheric conditions.

This classification supports a more robust interpretation of model–observation comparisons by explicitly accounting for the spatial representativeness of \textit{in situ} measurements relative to the model grid and satellite retrieval footprints. It also enables stratified analysis of model bias and data assimilation impact across different chemical regimes and land use types. AirNow hourly \textit{in situ} measurements of $\mathrm{NO}_2$ and ozone, along with site classification information, were downloaded from the public S3 bucket at \url{https://s3-us-west-1.amazonaws.com/files.airnowtech.org/airnow/YYYY/YYYYMMDD/HourlyData_YYYYMMDDHH.dat} for August 2023. AirNow data were used in monitoring mode and only for model validation. %Figure~SM~\ref{fig:airnow_sites} shows the locations of AirNow sites, colored based on their classification.

\subsubsection{Pandora}
\label{subsubsec:pandora}

The Pandora spectrometer network is a ground-based system of high-resolution UV–Vis spectrometers designed to measure atmospheric trace gases through direct-sun and sky-scanning observations. Originally developed by NASA and deployed as part of the Pandora Project and the Pandonia Global Network (PGN), the network provides high-accuracy column measurements of $\mathrm{NO}_2$, ozone, and other trace species. Pandora instruments are deployed globally, with a strong presence in North America and Europe, including colocations with satellite overpass paths and air quality monitoring stations to support validation of satellite retrievals and atmospheric chemistry models.

In this study, we focus on direct-sun $\mathrm{NO}_2$ retrievals from the Pandora network, which yield high-precision total column $\mathrm{NO}_2$ with a typical uncertainty of 0.01–0.05~DU and high temporal resolution (typically 1–2 minutes under clear-sky conditions). The $\mathrm{NO}_2$ SCDs are retrieved using DOAS in the 400–460~nm spectral window. These SCDs are converted to VCDs using geometric AMFs \citep{herman_no_2009}. As such, Pandora $\mathrm{NO}_2$ retrievals offer an important ground-based constraint with high representativeness for local conditions, especially in source regions with strong spatial and temporal gradients. Pandora observations are particularly valuable for validating and complementing satellite $\mathrm{NO}_2$ products, such as those from TROPOMI, and for evaluating model simulations and data assimilation output.

In this study, total column $\mathrm{NO}_2$ measurements from the Pandora network over CONUS serve as an independent reference for assessing the accuracy of $\mathrm{NO}_2$ fields at high temporal resolution and for quantifying potential representativeness errors in both model and satellite datasets. The air quality classification for Pandora sites was performed using gridded population density information \citep{gao_mapping_2020}. The sum of low-, medium-, and high-density values for each grid cell was calculated, and using arbitrary thresholds, rural, suburban, and urban classifications were assigned to each station.
% more info in SM

\subsubsection{Aircraft field campaigns}
\label{subsubsec:aircraft}

The Atmospheric Emissions and Reactions Observed from Megacities to Marine Areas (AEROMMA) is a multi-platform field campaign led by the National Oceanic and Atmospheric Administration (NOAA), aimed at improving the understanding of urban and marine atmospheric composition, emissions, and chemical processes. Conducted during the summer of 2023, AEROMMA integrated airborne, ground-based, and satellite observations to characterize the evolution of pollutants such as nitrogen oxides (\ce{NOx}), ozone, volatile organic compounds (VOCs), and particulate matter across diverse environments, including major U.S. cities and adjacent coastal and marine regions. A central objective of AEROMMA is to refine emission inventories, improve chemical transport models, and support satellite validation and air quality forecasting through comprehensive observations and coordinated data assimilation efforts \citep{noaa-csl_aeromma_nodate}.

The Synergistic TEMPO Air Quality Science (STAQS) field campaign, led by NASA, is a key component of the broader validation and science support effort for the TEMPO mission. Conducted in the summer of 2023, STAQS was designed to provide comprehensive ground-based and airborne observations to evaluate and enhance the interpretation of geostationary air quality measurements from TEMPO. The campaign targeted key urban regions in the southeastern and northeastern United States, including areas with complex emissions and photochemistry such as Atlanta and New York City. STAQS integrated multiple observing platforms, including NASA aircraft, ground-based remote sensing (e.g., Pandora spectrometers), \textit{in situ} surface monitors, and mobile laboratories, to characterize spatial and temporal variations in $\mathrm{NO}_2$, ozone, and other reactive trace gases at scales relevant to satellite pixels \citep{nasa_staqs_nodate}.

In situ measurements of $\mathrm{NO}_2$ onboard the NASA DC-8 aircraft, and column $\mathrm{NO}_2$ measurements from the Geostationary Coastal and Air Pollution Events Airborne Simulator (GCAS) instrument \citep{kowalewski_remote_2014}, deployed aboard the NASA Gulfstream V (GV), were used in this work. 
%TODO: Add reference for aeromma instrument

\subsection{Description of the JEDI system}
\label{subsec:description_jedi}

JEDI (Joint Effort for Data assimilation Integration) is a cutting-edge data assimilation (DA) framework built using object-oriented and generic programming principles. Its modular design promotes the separation of concerns, allowing scientists with expertise in different data assimilation components to collaborate effectively and in parallel without requiring a deep understanding of the entire system.

At the core of JEDI is OOPS (Object-Oriented Prediction System), which implements DA algorithms using abstract interfaces for key data assimilation components such as forecast models, observation operators, and background error covariance matrices. This abstraction makes the system versatile, enabling seamless integration of different forecasting models and observation types while minimizing code duplication. OOPS supports various variational, ensemble-based, and hybrid DA algorithms. Users can select their specific configurations for each assimilation component through the use of \textit{Yet Another Markup Language} (YAML) configuration files. Below, we describe each component and the configuration used in this study.

\subsubsection{Observation operators}
\label{subsubsec:obs_operators}

The Unified Forward Operator (UFO) is a central component of the JEDI framework responsible for simulating observations from model state variables, enabling the comparison between model forecasts and observations during data assimilation. UFO provides a standardized and extensible interface for many observation types, including satellite radiances using radiative transfer models, satellite retrievals, \textit{in situ} measurements, etc. UFO also handles filtering, quality control, bias correction, and the computation of observation-minus-background (OMB) and observation-minus-analysis (OMA) statistics.

UFO is model agnostic and accepts extracted model state information or Geophysical Variables at Locations (GeoVaLs) as input. GeoVaLs are sets of model background variables interpolated to the observation location and time that the observation operator requires. Each model interface is responsible for calculating GeoVaLs, and the OOPS component orchestrates the flow of information from the model interface to UFO. Once constructed, the GeoVaLs are passed to UFO, where the specific observation operator handles any additional vertical processing, such as vertical interpolation, coordinate transformations, or integration, based on the observation’s characteristics. This modular design ensures that interpolation is handled consistently across different models, while observation-specific logic remains encapsulated within each operator, supporting flexibility, code reusability, and scalability across the JEDI system. The two main observation operators used in this study are described below.

\paragraph{Column retrieval operator}

The column retrieval operator in UFO is a generic observation operator designed to assimilate vertically integrated retrievals from satellite, airborne, and ground-based instruments. It enables a consistent two-step assimilation approach by incorporating averaging kernels and \textit{a priori} profiles provided with retrieval products to map model state (i.e., trace gas mixing ratios) to observation space (i.e., tropospheric or total column values). In this study, we apply the column retrieval operator across retrievals from TROPOMI, TEMPO, GCAS, and Pandora, which demonstrates its flexibility in handling data from various platforms with varying spatial resolution, temporal coverage, and retrieval methodology. 

We focus here on the assimilation of $\mathrm{NO}_2$ retrievals, but the operator is also fully capable of handling other species and has been tested with products for carbon monoxide (CO) and ozone, among others. Its configuration-driven design allows users to tailor observation-specific YAML parameters without the need to modify the code, and its tangent-linear and adjoint parts allow for integration into variational and hybrid data assimilation systems. This generality supports the consistent assimilation of diverse observational datasets within JEDI. In this study, we use this operator with the TEMPO, TROPOMI, and GCAS tropospheric column $\mathrm{NO}_2$ observations, and Pandora total column $\mathrm{NO}_2$ observation.

More detailed documentation about the column retrieval operator is available on the JEDI documentation page: \url{https://jointcenterforsatellitedataassimilation-jedi-docs.readthedocs-hosted.com/en/latest/inside/jedi-components/ufo/obsops.html#column-retrieval-operator}

\paragraph{Vertical interpolation operator}

The vertical interpolation operator in UFO maps model fields from their native vertical coordinate system to the vertical locations of observations. This operator is useful for observations that report values at specific altitudes or pressure levels, such as surface monitoring networks and airborne \textit{in situ} measurements. This study uses the vertical interpolation operator to assimilate surface-level data from AirNow and aircraft-based observations from AEROMMA. The operator receives horizontally interpolated model fields via GeoVaLs and performs vertical interpolation within the observation operator itself, enabling the accurate extraction of model values at the vertical position of each observation. It supports linear and log-pressure interpolation methods, depending on the variable and coordinate system. The tangent linear and adjoint are trivial in this case and therefore support variational implementations. This flexible operator is well-suited for assimilating \textit{in situ} atmospheric composition measurements and supports consistent treatment of vertically resolved observations across various platforms. Further technical details are available in the JEDI documentation: \url{https://jointcenterforsatellitedataassimilation-jedi-docs.readthedocs-hosted.com/en/latest/inside/jedi-components/ufo/obsops.html#vertical-interpolation}

\subsubsection{Background error covariance matrix}
\label{subsubsec:saber}

System-Agnostic Background Error Representation (SABER) is a core component of the JEDI system responsible for representing background error covariances in a modular and extensible manner. Designed to be model-agnostic, SABER allows users to configure a wide range of covariance structures, including static, ensemble-based, and hybrid formulations, through flexible YAML-based configuration files. It supports multivariate covariances and localization methods and integrates with variational and ensemble-based data assimilation algorithms within JEDI. SABER is structured around a composition of building blocks, including variable changes, interpolation operators, and covariance models combined at runtime to construct the background error term (\( \mathbf{B} \)) used in the cost function. This design enables reproducibility and cross-model consistency in background error specification across different applications supported by the JEDI framework.

Background error on an Unstructured Mesh Package (BUMP) is a key component within SABER that provides efficient and flexible tools for the specification and manipulation of background error covariance models in JEDI. BUMP enables the construction of spatially and vertically varying covariance structures through diagnostics computed from ensemble statistics or static estimates. It supports various localization functions, spectral filtering techniques, and balance operators to model covariances in univariate and multivariate contexts. BUMP operates on structured grids and is optimized for high-performance computing environments, making it suitable for large-scale applications. BUMP thus serves as the computational backbone for many SABER-based covariance configurations in JEDI, promoting consistency and flexibility across different models and assimilation systems.

In this study, we employed a full-ensemble background error covariance \(\mathbf{B}\) formulation, in which no static component is included; all variance and covariance structures are derived solely from the ensemble. Localization was applied using the BUMP functionality called Normalized Interpolated Convolution on an Adaptive Subgrid (NICAS) \citep{menetrier_normalized_2020}. We implemented a horizontally varying localization radius, increasing from 100~km near the surface to 5000~km at the top of the atmosphere. In the vertical, a uniform localization of 0.3 was applied in the eta coordinate system (\href{https://glossary.ametsoc.org/wiki/Eta_vertical_coordinate}{AMS Glossary}). While this localization configuration, particularly in the vertical, could benefit from further refinement, it was deemed sufficient for this system demonstration.

\subsubsection{Data Assimilation Algorithm}
\label{subsubsec:da_algo}

OOPS is the data assimilation engine within the JEDI framework, responsible for implementing and managing optimization algorithms in a modular, model-agnostic environment. It serves as the central orchestrator that combines key components of the assimilation system, including the forecast model interface, background error representations (e.g., SABER), observation operators (via UFO), and minimization routines. Among the algorithms supported by OOPS is 4DEnVar (Four-Dimensional Ensemble Variational assimilation), a hybrid method that combines the strengths of ensemble and variational approaches. In the 4DEnVar formulation \citep{kleist_osse-based_2015-1}, background error covariances are estimated from a time-evolving ensemble of short forecasts, allowing flow-dependent, multivariate error structures. Unlike traditional 4D-Var, 4DEnVar avoids the need for tangent-linear and adjoint model integrations by projecting the cost function into the ensemble subspace. This makes it computationally efficient and more accessible for high-resolution or complex models.

A key strength of 4DEnVar is its ability to assimilate observations distributed in time over an assimilation window, making it particularly well-suited for incorporating geostationary observations such as those from TEMPO. Geostationary instruments provide hourly or sub-hourly retrievals, offering detailed information on the diurnal evolution of atmospheric composition. While 4DEnVar does not assimilate observations at their exact time, it uses a time-interpolated representation of the ensemble trajectories to account for temporal variation within the assimilation window. This allows the algorithm to effectively incorporate the high temporal resolution of geostationary datasets, improving the characterization of rapidly evolving processes such as pollution transport, boundary layer dynamics, and photochemical activity. These capabilities make 4DEnVar a powerful approach for leveraging the full potential of next-generation atmospheric composition observations such as TEMPO $\mathrm{NO}_2$ retrievals in JEDI.

\subsection{FV3 model interface and the GEOS-CF forecast model}
\label{subsec:fv3}

The FV3-JEDI interface connects the Finite-Volume Cubed-Sphere (FV3) dynamical core to the JEDI framework, enabling the use of FV3-based models in a wide range of assimilation applications, including atmospheric composition. It provides essential components to define the model geometry, manage state variables, and handle increments as required by variational and hybrid algorithms. The FV3-JEDI interface has been used for assimilation of Aerosol Optical Depth (AOD) in \citet{huang_jedi-based_2023, wei_prototype_2024}.

In this study, we extended and validated the FV3-JEDI interface to ensure full support for trace gas concentrations (e.g., $\mathrm{NO}_2$, ozone, CO) as state variables and to enable the assimilation of emissions fields as part of the model control vector (to be discussed in a following paper). These enhancements required careful coordination between model field definitions, variable transformations, and I/O handling to ensure compatibility with UFO and SABER components. The result is an interface supporting assimilation experiments targeting atmospheric composition. Additional technical details are available in the FV3-JEDI documentation: \url{https://jointcenterforsatellitedataassimilation-jedi-docs.readthedocs-hosted.com/en/latest/inside/jedi-components/fv3-jedi/index.html}

The GEOS-CF system is a high-resolution, global atmospheric composition modeling and forecasting framework developed by NASA’s Global Modeling and Assimilation Office (GMAO). Built on the GEOS Earth system model, GEOS-CF integrates detailed representations of tropospheric and stratospheric chemistry with meteorological processes using the FV3 dynamical core to produce near-real-time forecasts of trace gases and air pollutants, including $\mathrm{NO}_2$, ozone, CO, and fine particulate matter (PM\(_{2.5}\)) \citep{keller_description_2021}. It operates at approximately 25~km horizontal resolution with 72 vertical levels and includes fully coupled meteorology–chemistry interactions. In this study, we use anthropogenic emissions from the Community Emissions Data System (CEDS) \citep{hoesly_historical_2018} in conjunction with the Harmonized Emissions Component (HEMCO) \citep{keller_hemco_2014,lin_harmonized_2021} preprocessor to provide consistent and spatially resolved emissions inputs, supporting realistic simulation of atmospheric composition.

GEOS-CF is driven by assimilated meteorological fields from the GEOS Forward Processing (GEOS-FP, \url{https://gmao.gsfc.nasa.gov/GMAO_products/NRT_products.php}), also known as ‘replay mode’ \citep{orbe_large-scale_2017}, a configuration in which the model is driven by archived meteorological fields produced through data assimilation in the GEOS-FP system. This approach allows us to prescribe realistic and dynamically consistent meteorology without the need to re-assimilate meteorological observations, significantly reducing computational cost while preserving the fidelity of atmospheric transport and dynamics. This setup is particularly well-suited for ensemble-based data assimilation experiments in atmospheric composition, as it ensures that transport processes are constrained by high-quality meteorological analyses and their associated ensembles. In this configuration, the perturbed meteorological and dynamical fields from the GEOS-FP ensemble can be directly reused to represent uncertainties in transport and meteorology within the composition assimilation system.

\subsection{The Ensemble of Data Assimilations (EDA) and data assimilation suite design}
\label{subsec:eda}

The Ensemble of Data Assimilations (EDA) is a technique designed to better characterize the posterior background error covariance. It estimates analysis uncertainty by executing an ensemble of independent data assimilation cycles, each incorporating perturbed inputs. Unlike traditional ensemble forecast systems, which typically generate a single analysis followed by an ensemble of forecasts, the EDA directly samples both background and observational uncertainties, producing a distinct 4DEnVar analysis for each ensemble member. This approach enables a more realistic and dynamically consistent representation of background error statistics, significantly reducing the need for covariance inflation tuning. In some applications, such as atmospheric composition, EDA may even eliminate the need for inflation \citep{barre_emission_2019}. Below is the list of perturbations injected into the EDA system:

\textbf{Observation perturbations} are generated by adding random noise consistent with the reported observational errors, assuming a Gaussian distribution centered on the original observation value with variance defined by the observation error covariance. By ensuring each member sees a slightly different version of the observational dataset, this approach allows the ensemble to explore a range of plausible analysis states, supporting a more realistic estimation of background error covariances and improving the stability and spread of the ensemble-based assimilation system.

\textbf{Meteorological perturbations:} The GEOS-FP operational ensemble is produced by NASA GMAO. Using this ensemble ensures the meteorological diversity across ensemble members. Each member provides a distinct meteorological trajectory based on perturbed initial conditions and assimilation inputs. These meteorological fields are used to drive the GEOS-CF replay ensemble member simulations, ensuring that each ensemble member experiences different transport pathways and atmospheric dynamics. This variability is critical for representing transport-driven uncertainty in atmospheric composition and allows the ensemble to capture the flow-dependent nature of background errors.

\textbf{Emissions perturbations} are applied independently to each member’s anthropogenic emissions. We use a pseudo-random field generator based on a gamma distribution to create spatially correlated multiplicative perturbation fields. These fields are generated using a Gaussian random field modulated by a gamma distribution to ensure positive-definiteness and realistic spatial variability. The perturbed emissions maintain the spatial patterns of the base inventory while introducing ensemble spread in emission magnitudes, which is essential for representing uncertainties in source strength and distribution. Detail of the emission perturbations parameterization is detailed in Table \ref{tab:emission_perturbation_params}. This approach enables the ensemble to account for uncertainties in emissions alongside those from transport and observations, contributing to a more comprehensive representation of background error in the 4DEnVar assimilation system. The evolution of ensemble spread in this study is illustrated and discussed in Section~\ref{subsubsec:ens_spread}.

\begin{table}[h!]
\centering
\resizebox{\textwidth}{!}{%
\begin{tabular}{|l|c|c|c|c|c|c|c|c|}
\hline
\textbf{Sector} & Agriculture & Energy production & Industry & Residential, Commercial and Others & Ships & Solvents & Traffic & Waste \\
\hline
\textbf{Perturbations magnitude} & 75\% & 30\% & 30\% & 45\% & 40\% & 50\% & 35\% & 55\% \\
\hline
\textbf{Spatial Correlation length scale} & 500km & 200km & 200km & 200km & 500km & 200km & 200km & 200km \\
\hline
\textbf{Temporal correlation} & 1 month & 1 hour & 1 hour & 1 hour & 1 month & 1 month & 1 hour & 1 month \\
\hline
\end{tabular}%
}
\caption{Sector-specific parameters used for emission perturbations in the ensemble system, including magnitude, spatial correlation scale, and temporal correlation.}
\label{tab:emission_perturbation_params}
\end{table}

\section{Results}
\label{sec:results}

\subsection{Observation-space analysis statistics}
\label{subsec:obs_space_ana}

A critical first step in evaluating any data assimilation system is to assess its performance in the observation space. This involves verifying that the assimilation effectively reduces the analysis error with respect to the assimilated observations. By comparing model-simulated equivalents of observations to the actual measurements, we can determine whether the assimilation is fulfilling its primary role of improving the fit to observed data before extending the evaluation to independent observations or forecast skill metrics.

To evaluate the impact of data assimilation on forecast quality, we begin by comparing the observations with the model-simulated equivalents, \( \mathbf{H}(\mathbf{x}) \), derived from both the background (forecast) and the analysis (assimilated) fields. A fit-to-observation performance metric is defined as the difference of the absolute errors between observation-minus-background (OMB) and observation-minus-analysis (OMA). Positive values of this metric indicate that the assimilation has improved agreement with the observations (i.e., OMA is smaller than OMB), bringing the analysis closer to the assimilated data.

\begin{figure}
    \centering
    \includegraphics[width=0.5\linewidth]{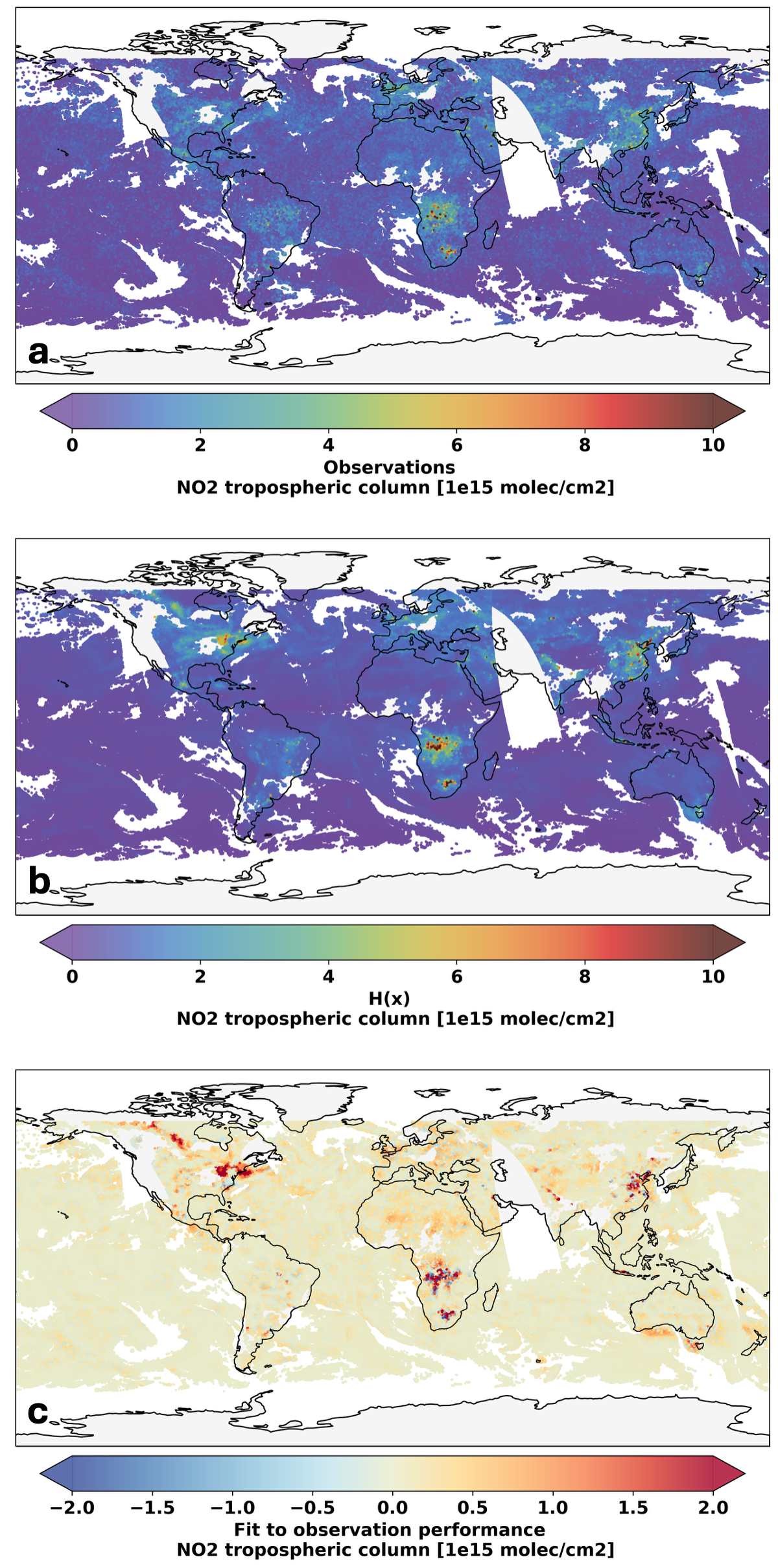}
    \caption{Tropospheric NO2 column on 9 August 2023, a) measured by TROPOMI, b) the model equivalent values on backgrounds or H(xb), and c) the fit to observation performance (|OMB| - |OMA|)}
    \label{fig:tropomi_no2_obs_hx_fit}
\end{figure}

Figure~\ref{fig:tropomi_no2_obs_hx_fit} shows the tropospheric $\mathrm{NO}_2$ column values on 9 August 2023 as (a) measured by TROPOMI (observation) and (b) simulated by the model (background or \( \mathbf{H}(\mathbf{x}_b) \)) using the column retrieval observation operator described in Section~\ref{subsubsec:obs_operators}. The corresponding fit-to-observation performance metric is presented in Figure~\ref{fig:tropomi_no2_obs_hx_fit}c. Predominantly positive values indicate that the assimilation of tropospheric $\mathrm{NO}_2$ column data generally improved the agreement with the TROPOMI retrievals. This initial result serves as a diagnostic check, confirming that the global data assimilation system is functioning as expected. The primary objective of this study, however, is to demonstrate a novel data assimilation capability for TEMPO $\mathrm{NO}_2$ retrievals.

\begin{figure}
    \centering
    \includegraphics[width=0.75\linewidth]{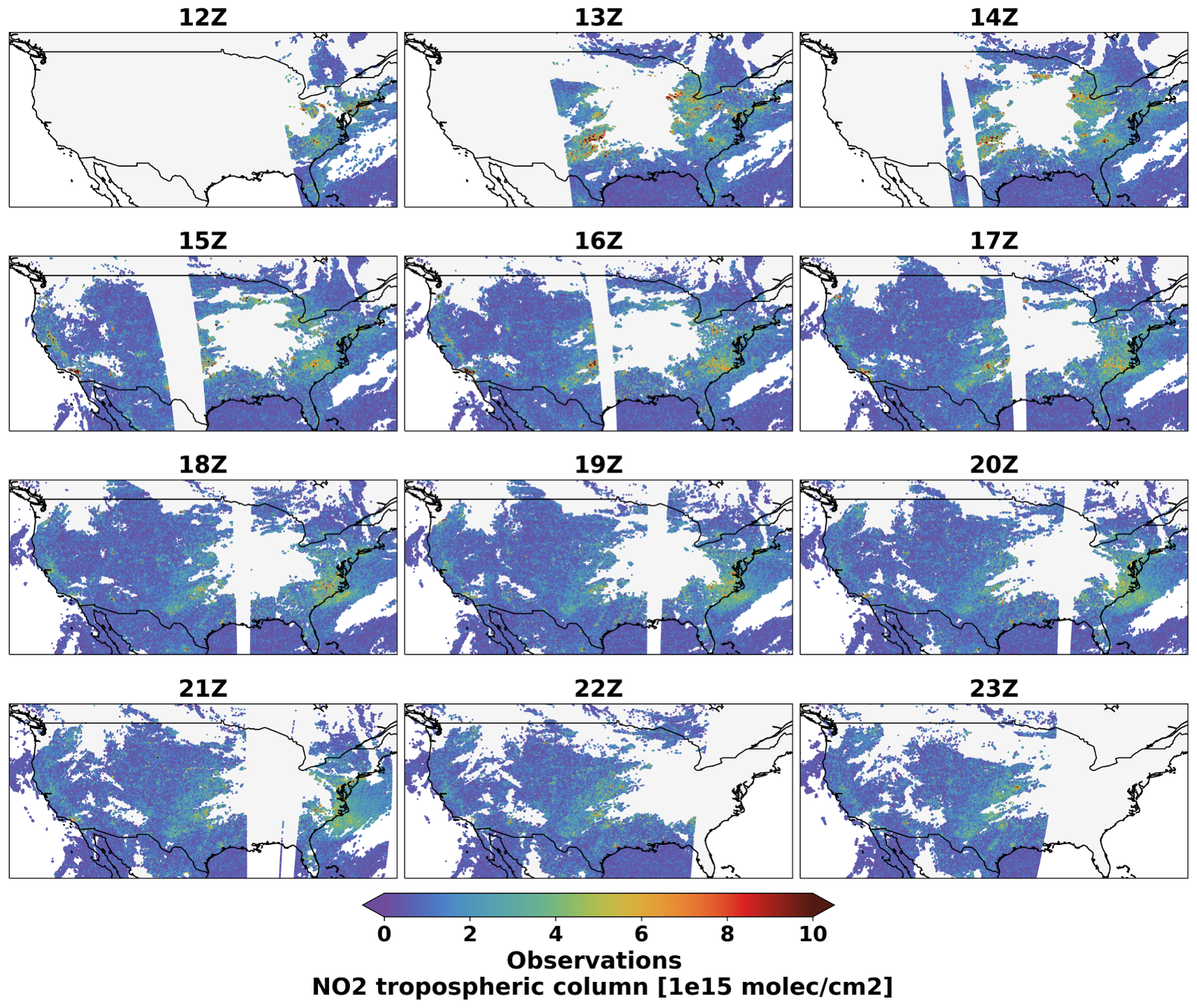}
    \caption{Hourly binned tropospheric NO2 column measured by TEMPO on 9 August 2023}
    \label{fig:tempo_no2_obs_hourly}
\end{figure}

\begin{figure}
    \centering
    \includegraphics[width=0.75\linewidth]{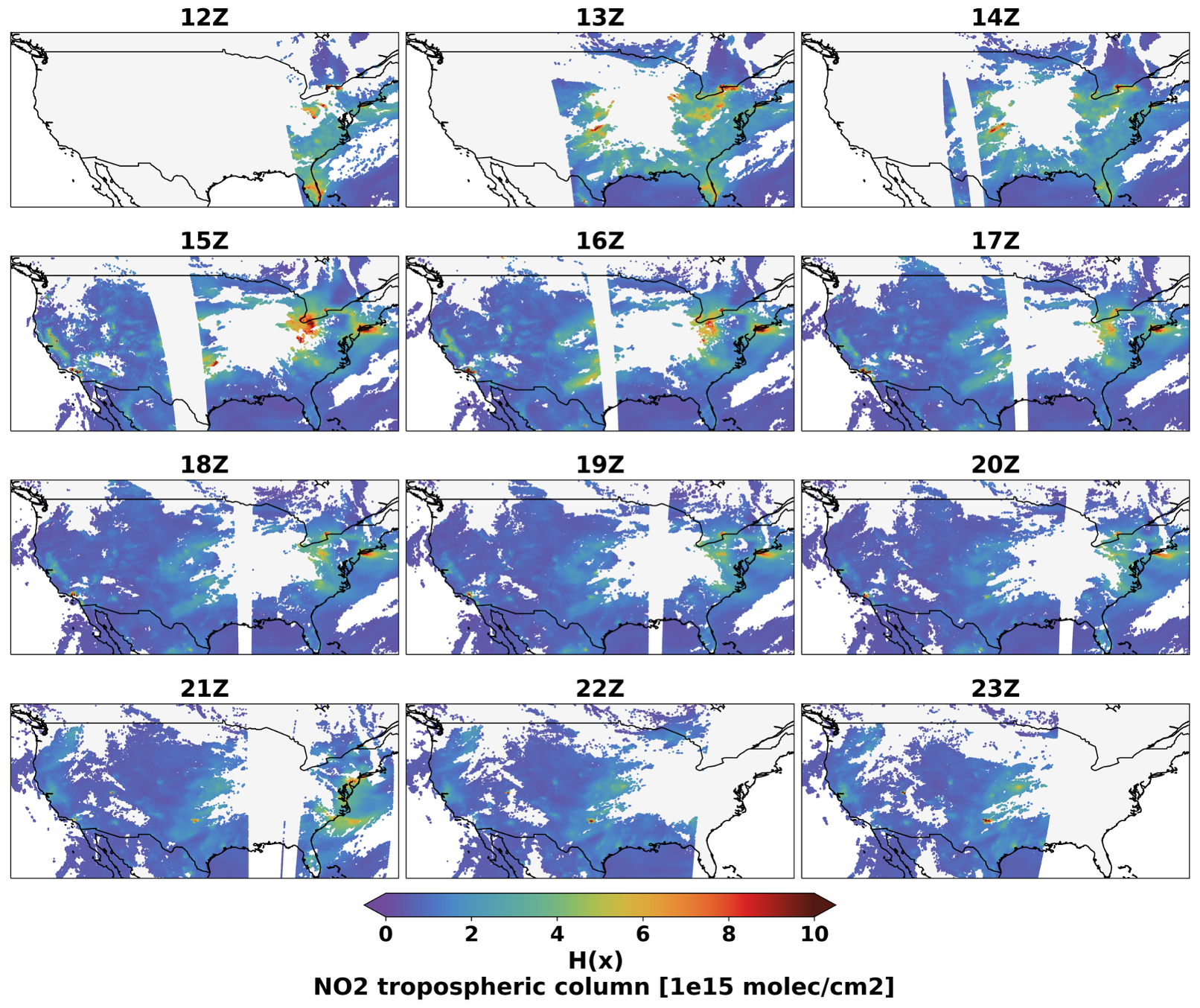}
    \caption{Hourly model equivalent background tropospheric NO2 column or H(xb) on 9 August 2023}
    \label{fig:tempo_no2_hxb_hourly}
\end{figure}

\begin{figure}
    \centering
    \includegraphics[width=0.75\linewidth]{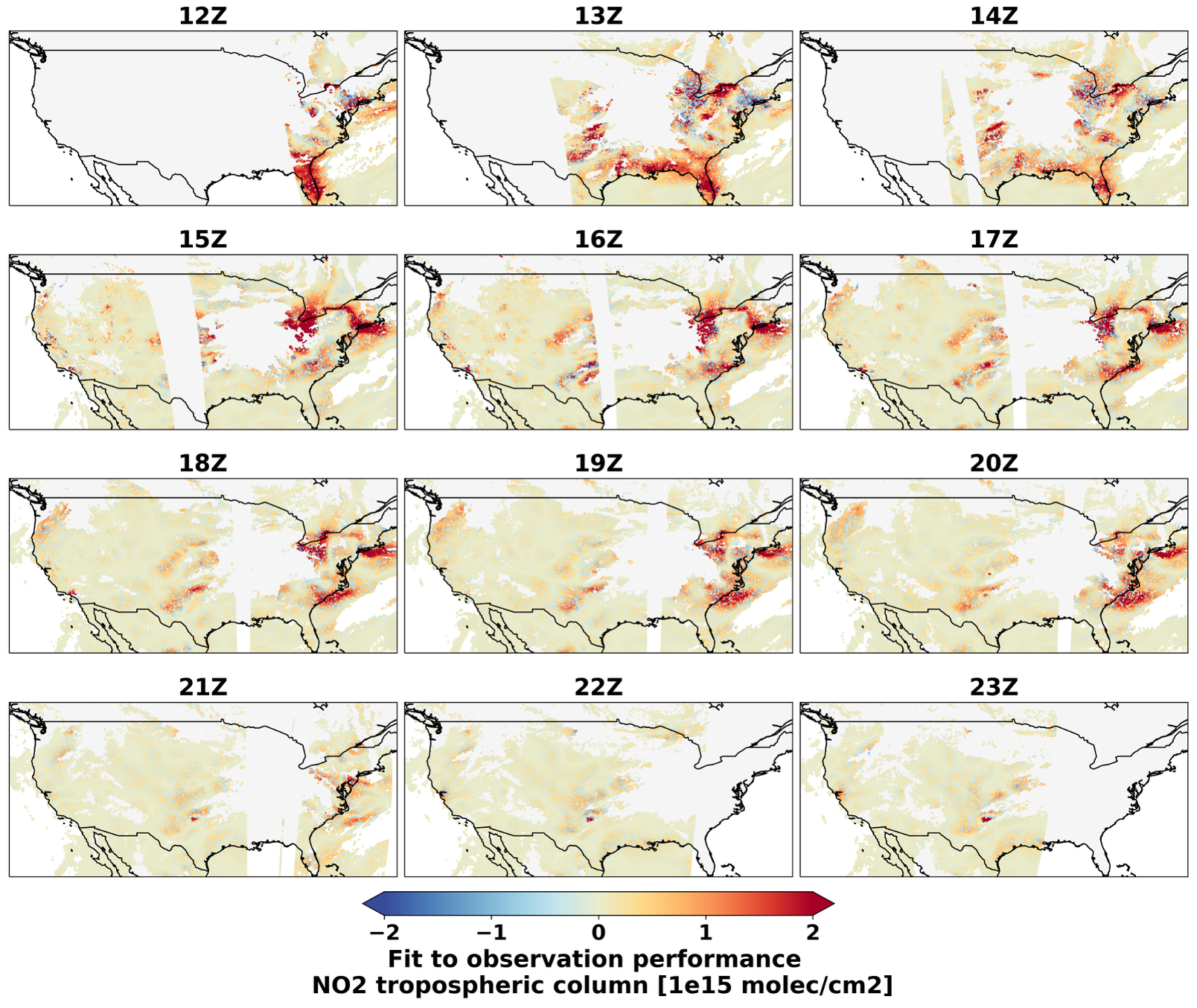}
    \caption{Hourly fit to observation performance (|OMB| - |OMA|) on 9 August 2023}
    \label{fig:tempo_no2_fit_hourly}
\end{figure}

Figures~\ref{fig:tempo_no2_obs_hourly},~\ref{fig:tempo_no2_hxb_hourly}, and~\ref{fig:tempo_no2_fit_hourly} illustrate the hourly tropospheric $\mathrm{NO}_2$ column observations from TEMPO over CONUS on 9 August 2023, alongside the corresponding model values, \(\mathbf{H}(\mathbf{x})\), and the fit-to-observation performance, respectively. Since TEMPO covers only North America, in this study, we focus on the CONUS domain when assessing the impact of TEMPO assimilation. Similar to TROPOMI, the model equivalents were computed using the column retrieval observation operator applied at the time of observation. The fit-to-observation metric shows both spatial and temporal variability, highlighting the advantage of using a 4D assimilation framework.

In this case study over CONUS (Figure~\ref{fig:tempo_no2_fit_hourly}), the most pronounced error reductions are observed during the early morning hours, particularly between 12Z and 14Z, with smaller error reductions in the afternoon. The morning improvements coincide with the peak in anthropogenic $\mathrm{NO}_2$ emissions, primarily driven by increased traffic activity during the morning rush hour, as well as enhanced photochemical activity due to rising solar radiation. Planetary Boundary Layer (PBL) growth in the morning and overestimation of nighttime $\mathrm{NO}_2$ concentrations can also contribute to the stronger early morning increment. The assimilation system effectively captures and corrects the elevated $\mathrm{NO}_2$ concentrations, demonstrating its sensitivity to diurnally varying emission, transport, and chemistry signals.

During the late afternoon, the overall impact of assimilation diminishes across most regions. This reduction is consistent with a natural decline in $\mathrm{NO}_2$ emissions and photochemical production toward the evening. By this time, $\mathrm{NO}_2$ concentrations tend to be lower and more homogeneous, reducing the magnitude of analysis increments required. In the Southeastern United States, the correction is not evident during the late morning and afternoon (15Z to 23Z), suggesting that the information introduced by assimilation earlier in the day is successfully propagated forward through the cycling of the data assimilation system. This indicates that the analysis increments applied in the morning are retained and influence subsequent background states, reducing the need for additional corrections later in the day.

In other regions, such as the Long Island study region, the assimilation corrections occur later in the day (15Z to 20Z). This pattern likely reflects differences in local emission profiles, transport dynamics, and chemical regimes, where $\mathrm{NO}_2$ enhancements are less sharply peaked and more persistent.

Overall, these results demonstrate the data assimilation system’s ability to dynamically adjust the model state in response to high-frequency, spatially resolved observations. The observed improvements in agreement with hourly TEMPO $\mathrm{NO}_2$ retrievals underscore the value of dense, geostationary observations in capturing the rapid temporal variability of $\mathrm{NO}_2$ chemistry.

\begin{figure}
    \centering
    \includegraphics[width=1\linewidth]{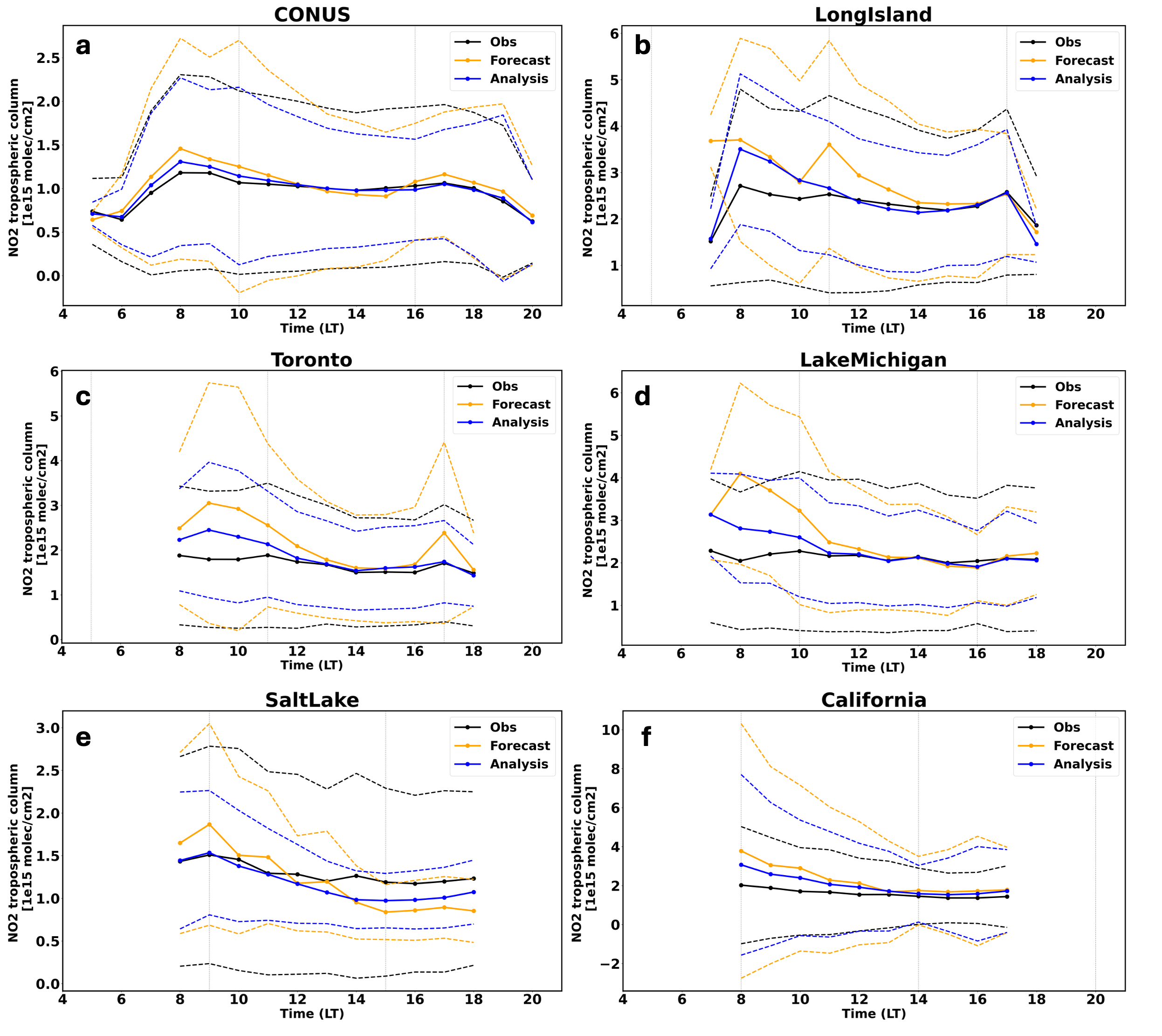}
    \caption{Mean diurnal values of tropospheric NO2 column averaged over (a) CONUS and (b-f) study regions. Observation values are in black, and the model-equivalent values of forecast and analysis are in blue and orange, respectively. Dashed lines indicate ±1 standard deviation range from the mean.}
    \label{fig:tempo_obs_space_diel}
\end{figure}

Next, we examined the statistical diurnal variability of the tropospheric $\mathrm{NO}_2$ column. The goal here is to showcase and assess how the 4DEnVar assimilation method is suited to modulate the diurnal $\mathrm{NO}_2$ variability while enhancing the accuracy of the predicted fields. Figure~\ref{fig:tempo_obs_space_diel} shows the diurnal variation of tropospheric $\mathrm{NO}_2$ columns averaged over CONUS and over the study regions for the period of the experiment, i.e., 4 to 31 August 2023. Observations and model-equivalent values, \(\mathbf{H}(\mathbf{x})\) (both forecast and analysis), were binned by hour, and the mean and standard deviation were computed for each hour. The resulting diel cycle is plotted in local time (LT), with the central time zone used for CONUS. Dashed lines indicate the \(\pm1\) standard deviation range from the mean, and vertical grey lines mark the 6-hourly data assimilation window bounds.

In Figure~\ref{fig:tempo_obs_space_diel}, TEMPO observations reveal elevated $\mathrm{NO}_2$ values around 8~a.m.~LT due to morning rush hour emissions and nighttime $\mathrm{NO}_2$ accumulation. This increase is also reflected in the forecast and analysis fields. However, discrepancies between the forecast and observed morning $\mathrm{NO}_2$ levels suggest potential errors in the model's representation of morning $\mathrm{NO}_2$ concentrations. Accurately capturing the morning $\mathrm{NO}_2$ buildup is particularly challenging due to emission timing and magnitude uncertainties, nighttime and early morning chemistry complexities, and difficulties in representing the evolving PBL. The larger standard deviation range, highlighting stronger variability in observations, analyses, and forecasts, corroborates this elevated uncertainty. Assimilating TEMPO’s early morning observations brings the analysis fields closer to the observed values. This improvement is consistent across CONUS and within all defined study regions. Overall, the analysis fields exhibit narrower standard deviation bands and better alignment with observations compared to the background, particularly during the morning hours. While biases in the afternoon are generally smaller, the assimilation still leads to improvements in the analysis fields throughout the day.

\subsection{Model-space analysis statistics}
\label{subsec:model_space_ana}

\subsubsection{Assimilation increments}
\label{subsubsec:assim_inc}

\begin{figure}
    \centering
    \includegraphics[width=0.5\linewidth]{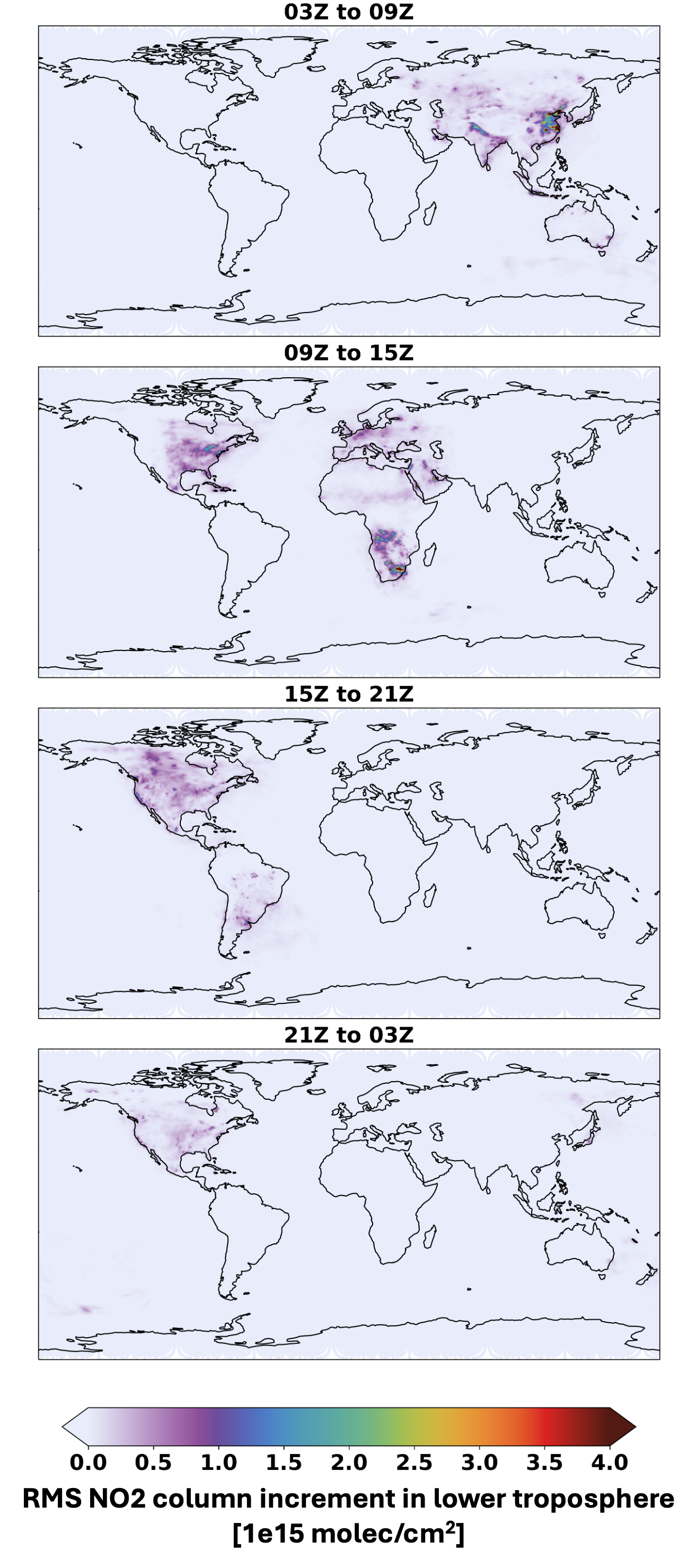}
    \caption{RMS of \ce{NO2} partial column increments within the lower tropospheric layer (surface to 800hPa) for each assimilation window during the experiment period}
    \label{fig:global_rms_inc}
\end{figure}

In this section, we examine the spatial and temporal characteristics of the DA increments (analysis-minus-background). As a first diagnostic, we evaluate the overall magnitude of the increments by computing the root mean square (RMS) of the differences between the analysis and background fields. Figure~\ref{fig:global_rms_inc} presents the RMS of $\mathrm{NO}_2$ increments, computed as partial column values within the lower troposphere (surface to 800~hPa), averaged over the full experimental period. Although analyses are produced hourly, RMS values are computed for each assimilation window to provide a consistent, global diagnostic across time.

The results show, as anticipated, that TROPOMI observations exert a once-daily influence, primarily outside the spatial domain observed by TEMPO. Meanwhile, the North American region, where TEMPO provides hourly coverage, receives adjustments during three out of four assimilation windows per day. This temporal pattern reflects the different observation schedules and viewing geometries of the two instruments. Additionally, the magnitude of the corrections introduced by both TEMPO and TROPOMI observations appears to be of similar order in the plots, despite the differing spatial coverage and revisit frequencies.

These diagnostics, in line with those presented in section~\ref{subsec:obs_space_ana}, provide an initial validation of the assimilation system’s behavior. They demonstrate that the global DA system is correctly integrating information from multiple satellite instruments with complementary characteristics. While these results highlight the potential for synergistic use of TEMPO and TROPOMI data and suggest possible bias interactions between the two datasets, such analysis is beyond the scope of this study. The primary goal here is to demonstrate the feasibility and performance of assimilating TEMPO $\mathrm{NO}_2$ retrievals within a global cycling system.

\begin{figure}
    \centering
    \includegraphics[width=1\linewidth]{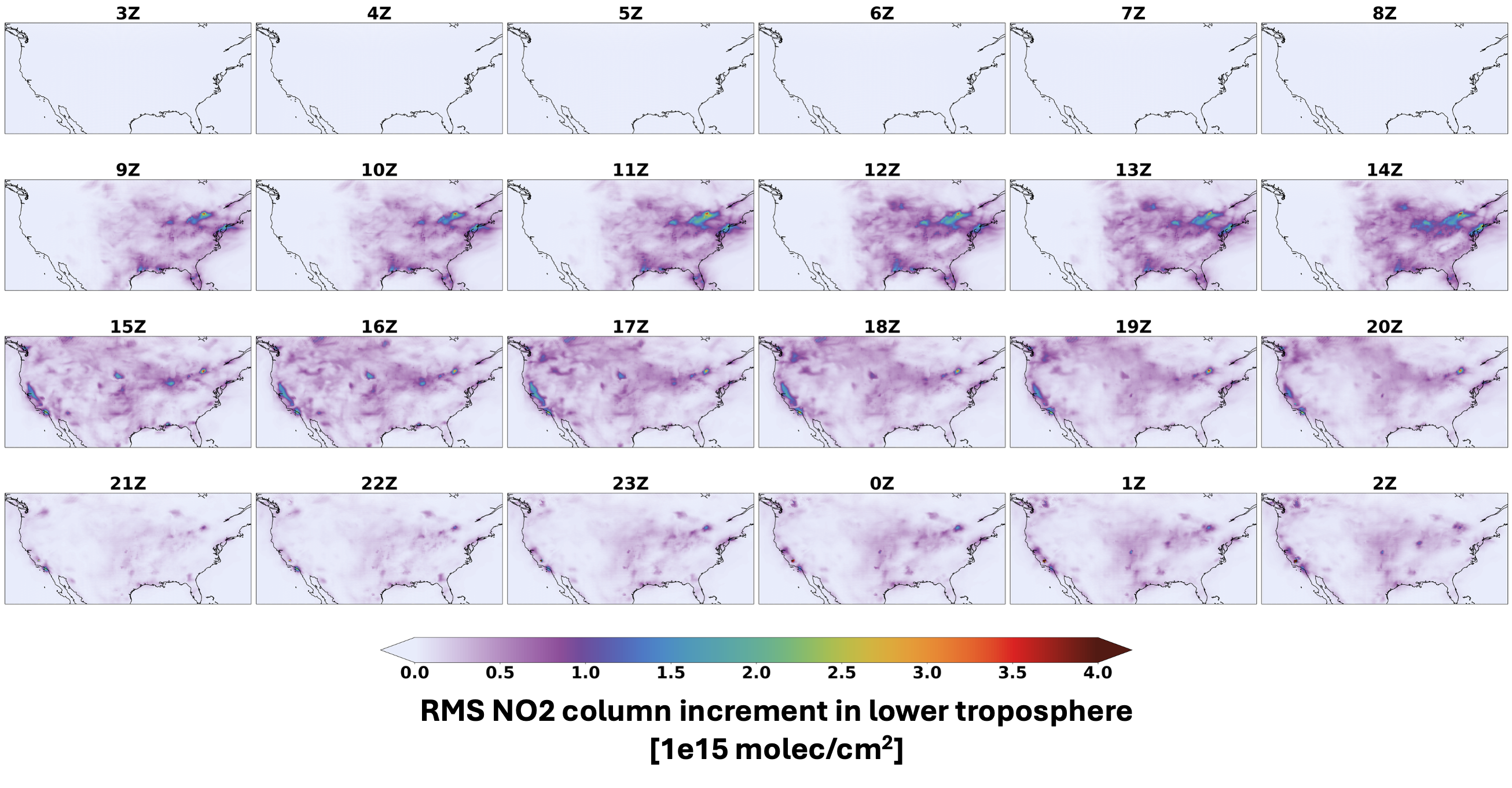}
    \caption{RMS of hourly \ce{NO2} partial column increments within the lower tropospheric layer (surface to 800hPa) for each assimilation window during the experiment period}
    \label{fig:conus_rms_inc_hourly}
\end{figure}

Figure~\ref{fig:conus_rms_inc_hourly} presents hourly $\mathrm{NO}_2$ analysis increments for each assimilation window, with a spatial focus on the CONUS region. Each row corresponds to one of the four assimilation windows in a 24-hour cycle, displaying the RMS of the increments derived from the analysis-minus-background fields over the experiment period.

As TEMPO $\mathrm{NO}_2$ retrievals are based on UV–Visible measurements, they are only available during daylight hours. Consequently, no TEMPO observations are assimilated during the 03–09Z window, which leads to zero analysis increments for this period due to the absence of observational input. The remaining three assimilation windows each contain TEMPO observations. The 09–15Z window typically includes morning observations beginning around 12Z; the 15–21Z window represents midday and generally contains continuous observational coverage throughout the window; and the 21–03Z window covers the evening period, with observations extending up to approximately 23Z.

Figure~\ref{fig:conus_rms_inc_hourly} demonstrates that the 4DEnVar assimilation method effectively propagates observational influence throughout the entire assimilation window by leveraging ensemble-based temporal correlations. As a result, analysis increments are not only evident at observation times but also during hours without direct observational input, such as 09–11Z and 00–02Z, highlighting the ability of the system to infer corrections based on temporally adjacent data.

\begin{figure}
    \centering
    \includegraphics[width=1\linewidth]{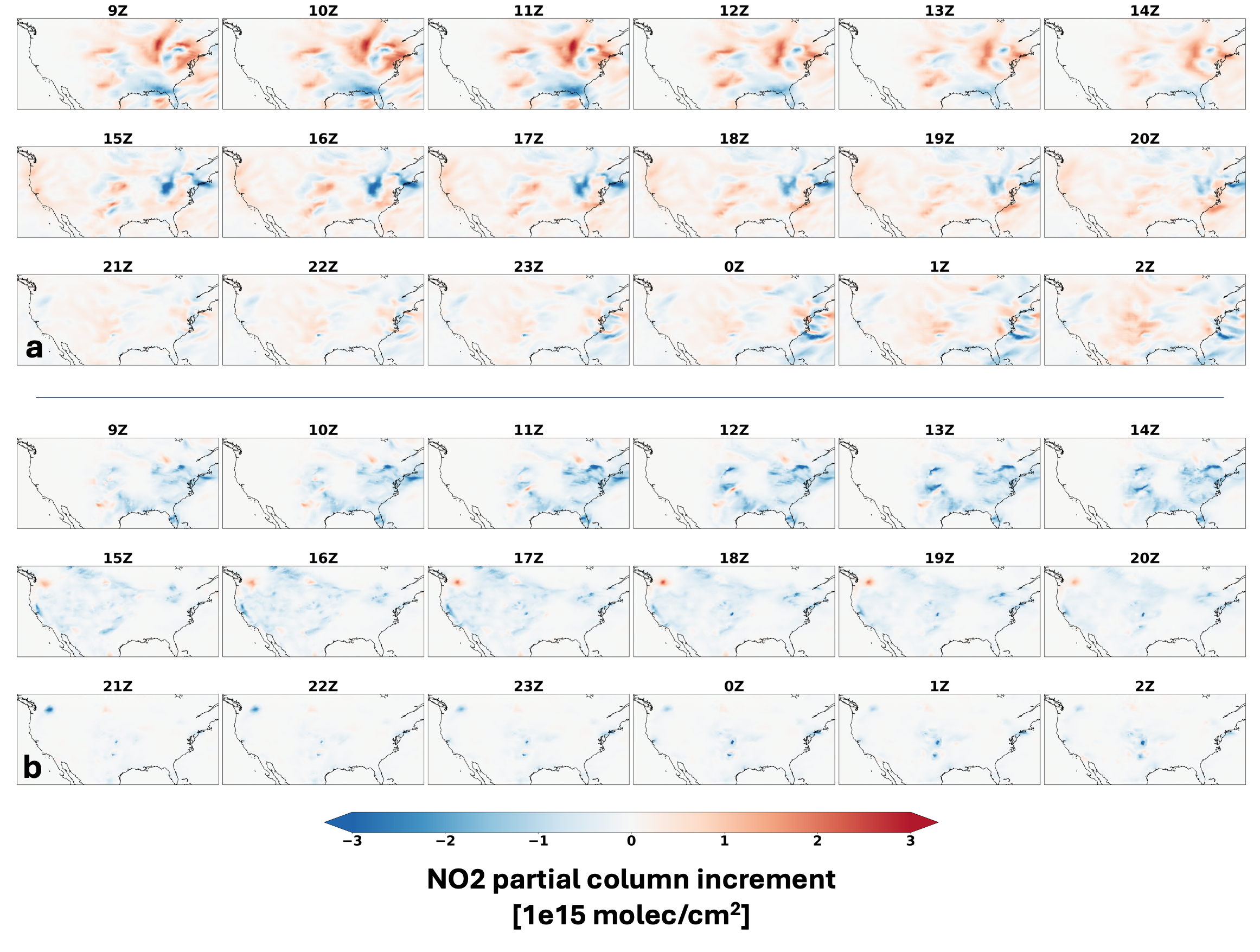}
    \caption{\ce{NO2} partial column increment within a) 800 hPa to TOA and b) surface to 800 hPa (free troposphere) on 9 August 2023. Hours with increment values of zero are excluded (3Z to 9Z)}
    \label{fig:conus_rms_inc_vertical_hourly}
\end{figure}

Next, we examine the vertical structure of the assimilation increments. Figure~\ref{fig:conus_rms_inc_vertical_hourly} presents a snapshot of hourly analysis increments of partial column $\mathrm{NO}_2$ on 9 August 2023, separated into two vertical layers: panel~(a), free troposphere to the top of the atmosphere (800~hPa to TOA), and panel~(b), lower troposphere (surface to 800~hPa). Within the lower troposphere (Figure~\ref{fig:conus_rms_inc_vertical_hourly}b), the analysis increments are predominantly negative, indicating that the background fields systematically overestimate near-surface $\mathrm{NO}_2$ concentrations relative to both TEMPO and TROPOMI observations. As noted in previous diagnostics (e.g., Figure~\ref{fig:tempo_obs_space_diel}), these negative corrections, reflecting a net removal of $\mathrm{NO}_2$, are a consistent feature of this experiment, especially during the morning hours. Although some localized regions exhibit positive increments, these are less common and typically occur during the early part of the diurnal cycle. The persistence of this negative increment pattern across multiple days will be discussed in more detail in the following sections.

In contrast, the structure of the increments in the upper layers (Figure~\ref{fig:conus_rms_inc_vertical_hourly}a) is less systematic and exhibits flow-dependent spatial variability. The snapshot in Figure~\ref{fig:conus_rms_inc_vertical_hourly}a shows a mixture of positive and negative increments aloft, suggesting adjustments linked to large-scale transport processes in the free troposphere and the upper troposphere–lower stratosphere (UTLS). Unlike the more uniform corrections near the surface, the variability in the upper-level increments points to a more complex vertical adjustment process and underscores the importance of assessing the performance of localization tuning in the 4DEnVar configuration.

Although the assimilation is conducted with a 32-member ensemble, the potential for spurious correlations remains a concern, particularly in regions or layers with limited observational constraint. Continued evaluation of localization settings will be necessary to optimize the accuracy and stability of upper-level corrections. Vertical localization plays a critical role in constraining how observational information is distributed across vertical levels in the assimilation system. A comparison between panels~(a) and~(b) of Figure~\ref{fig:conus_rms_inc_vertical_hourly} illustrates the resulting vertical structure of the analysis increments.

A similar temporal pattern is also evident in the upper layers, where the morning increments tend to be more pronounced. However, interpreting these results is complicated by the differing vertical thicknesses of the layers shown. Since the figures compare integrated quantities (molecules/cm\(^2\)) over layers of unequal depth, this could give the misleading impression that significant corrections are being applied inappropriately to upper layers during tropospheric column assimilation.

\begin{figure}
    \centering
    \includegraphics[width=0.75\linewidth]{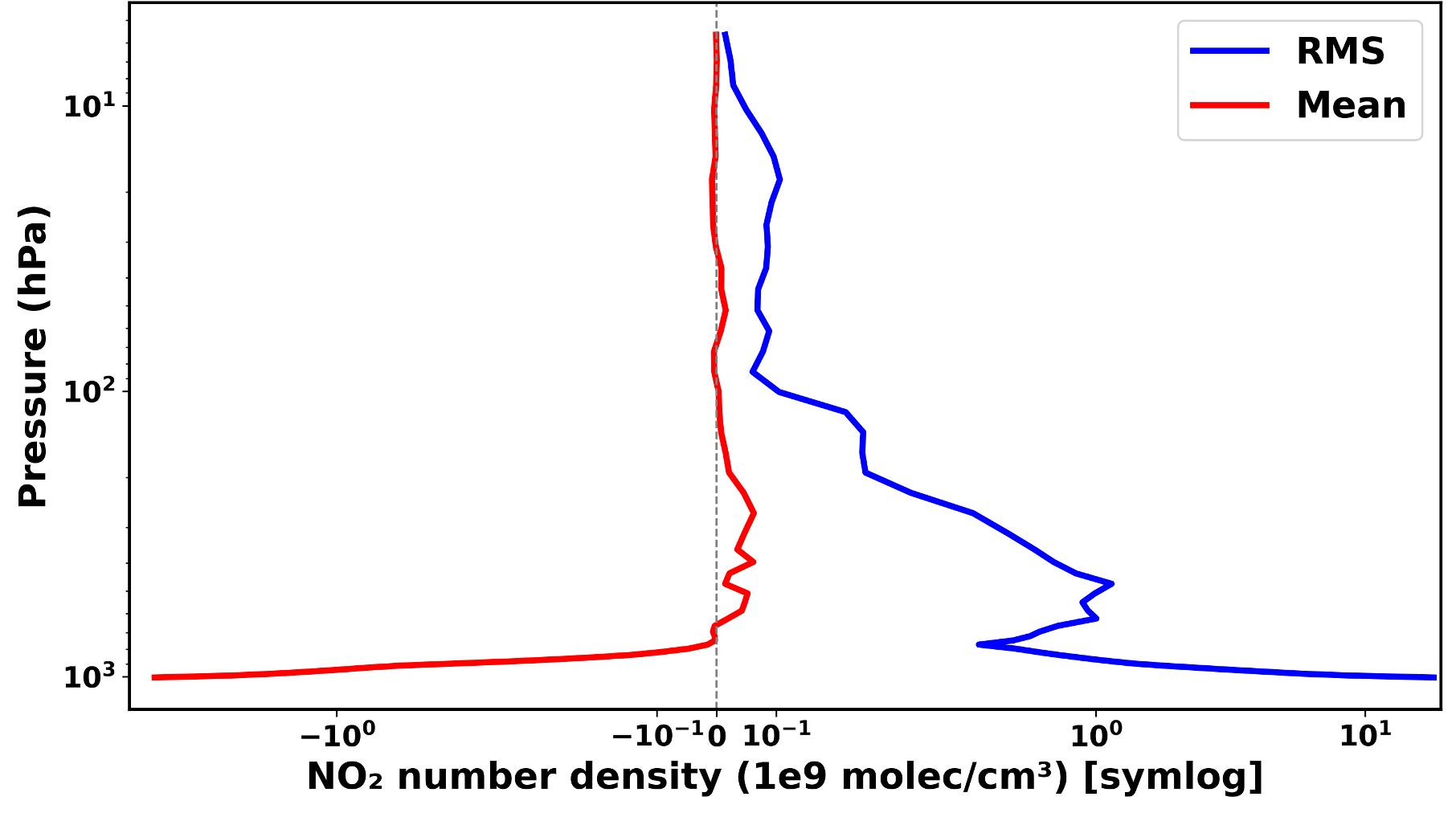}
    \caption{Vertical distribution of \ce{NO2} increment in number density unit of 1e9 molec/cm3 during 9 August 2023. The blue and red lines show the RMS and the mean of the increment, respectively. A log scale is used for the Y-axis, and a symlog scale with a threshold of 0.5 is used for the X-axis. }
    \label{fig:profile_increments}
\end{figure}

To address this, Figure~\ref{fig:profile_increments} presents vertical profiles of the mean and RMS $\mathrm{NO}_2$ increments in terms of number density (molecules/cm\(^3\)), offering a more physically consistent diagnostic of the vertical structure. These profiles confirm earlier findings: large, systematic negative increments are concentrated near the surface, while the free troposphere, UTLS, and lower stratosphere show substantial variability (RMS), but relatively small positive mean increments.

These diagnostics suggest a more complex vertical adjustment process, particularly aloft, and reinforce the need to carefully assess the performance of vertical localization. To explore this further, we conducted an additional test experiment from 4 August to 17 August 2023, using a more restrictive vertical localization configuration in the UTLS and stratosphere, effectively limiting the upward spread of observational impact beyond the upper troposphere. The results of this additional experiment (not shown) indicates that the near-surface corrections, both in magnitude and spatial structure, remain very similar under the modified localization, supporting the conclusion that the observed surface-level adjustments are well constrained by the available observations with the current broad vertical localization setup.

\subsubsection{Ensemble spread}
\label{subsubsec:ens_spread}

\begin{figure}
    \centering
    \includegraphics[width=0.75\linewidth]{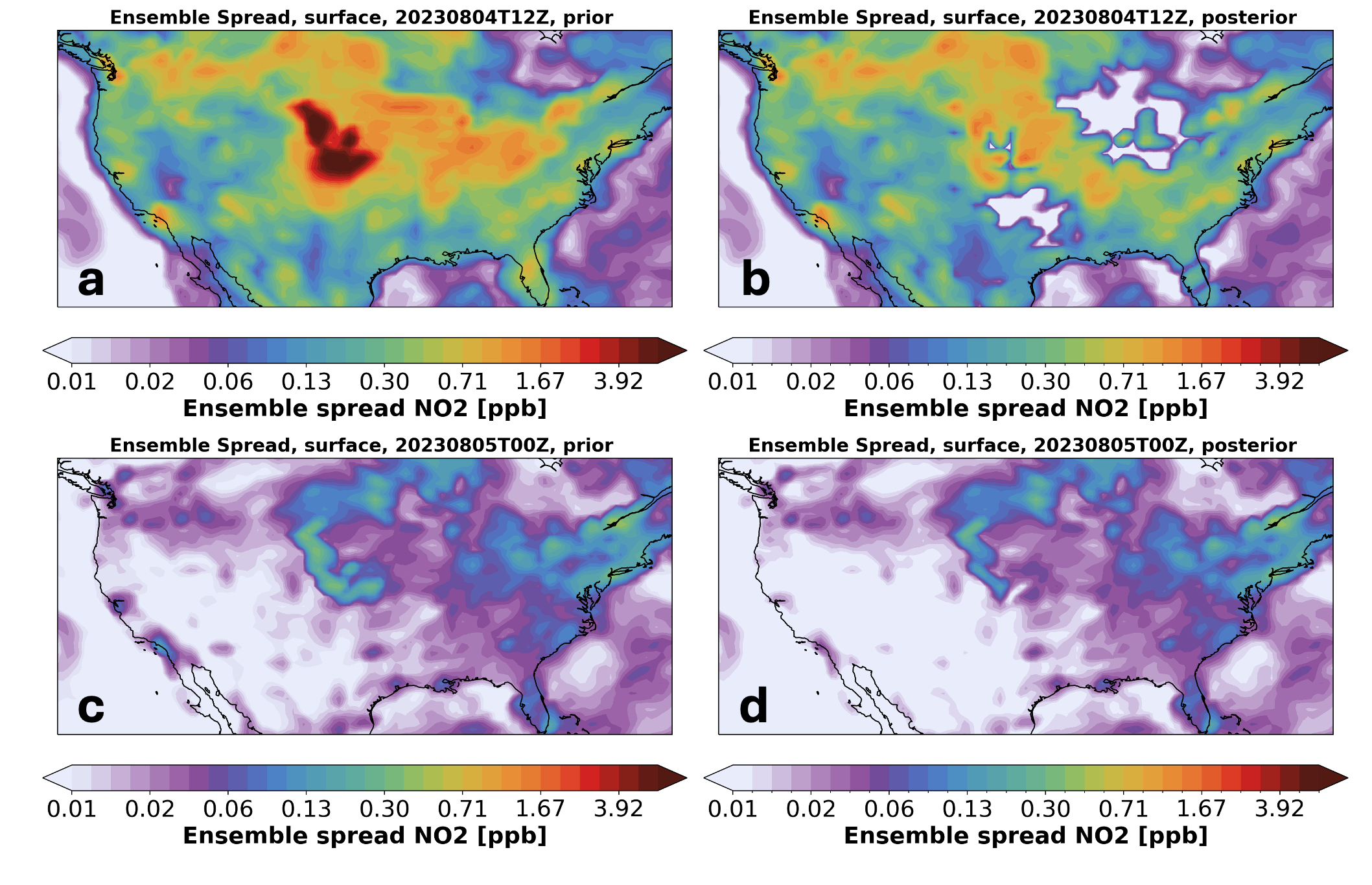}
    \caption{Ensemble spread of prior (left column) and posterior (right column) at a and b) 4 August 2023, 12Z, b, d)  5 August 2023, 00Z at surface}
    \label{fig:spread_surface_start}
\end{figure}

\begin{figure}
    \centering
    \includegraphics[width=0.75\linewidth]{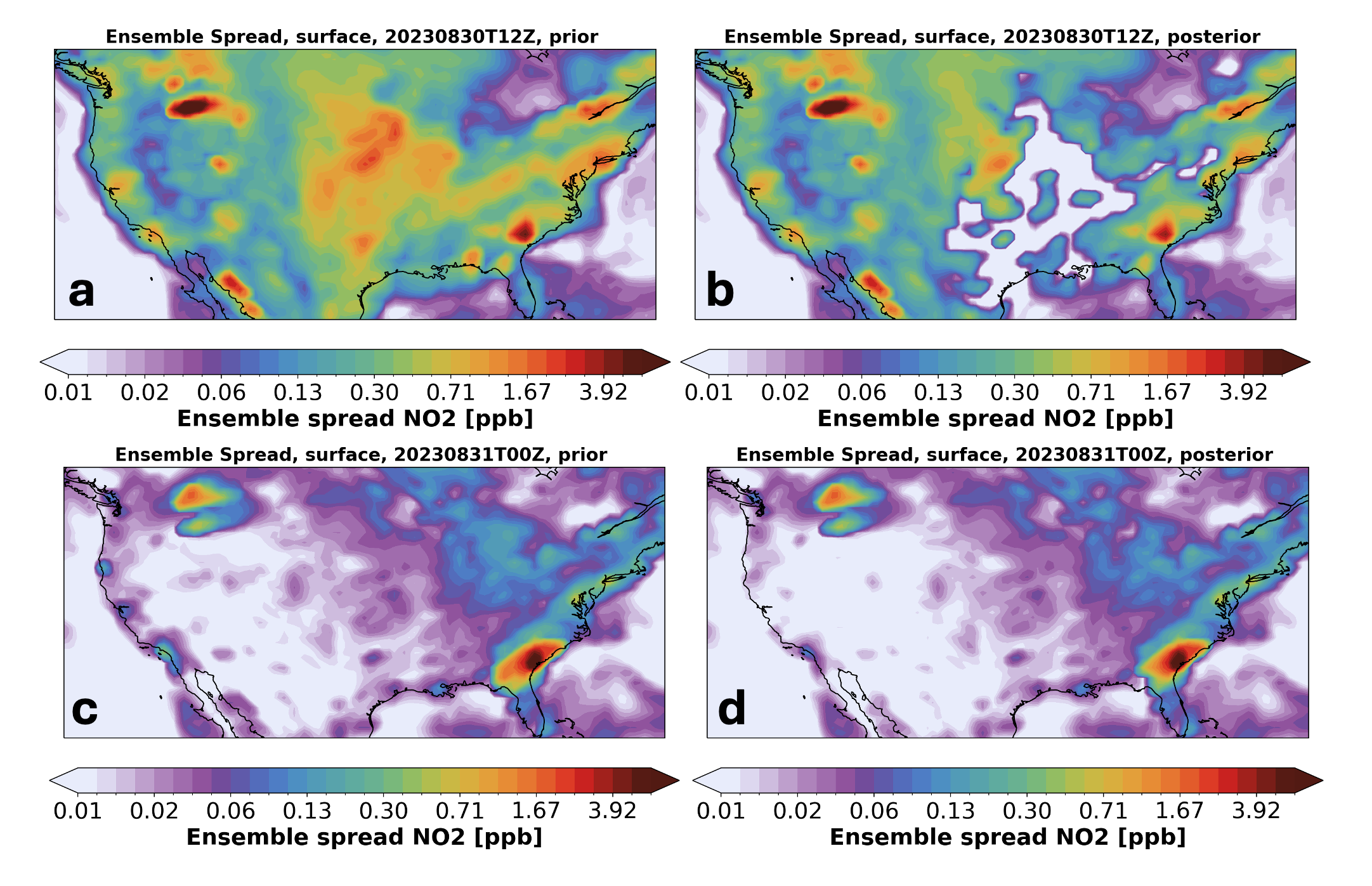}
    \caption{Ensemble spread of prior (left column) and posterior (right column) at a and b) 30 August 2023, 12Z, b, d)  31 August 2023, 00Z at surface}
    \label{fig:spread_surface_end}
\end{figure}

Figure~\ref{fig:spread_surface_start} shows surface $\mathrm{NO}_2$ ensemble spread (standard deviation) at 12Z and 00Z for the prior (left column) and posterior (right column) on 4 August 2023, a representative day at the beginning of the experiment. Figure~\ref{fig:spread_surface_end} shows similar quantities on 30 August 2023, a representative day at the end of the experiment. Contrasting the ensemble spread at 12Z and 00Z and on different days highlights the variability in hourly conditions, as well as the preservation of the ensemble spread during the experiment period.

Hourly changes in the ensemble spread show the system's sensitivity to the diurnal variability of $\mathrm{NO}_2$ concentrations. The spread levels in the troposphere and stratosphere have been checked as well, confirming continued maintenance of ensemble spread aloft, albeit with smaller magnitudes as expected (not shown). These reduced values are consistent with lower $\mathrm{NO}_2$ mixing ratios at higher altitudes and a diminished impact of emission perturbations in the upper and free troposphere. In the EDA framework, the cost function minimization is performed independently for each ensemble member, and the use of perturbed observations contributes to sustaining the ensemble spread over the course of the experiment. Comparison between posterior and prior ensemble spread of surface $\mathrm{NO}_2$ shows similar patterns but slightly smaller magnitudes for the posterior as the solutions of the analyses are converging towards the observational constraints, reducing the posterior error distribution, but do not collapse to a very small spread.

\subsection{Evaluations against independent observations}
\label{subsec:eval}

In this section, we evaluate the performance of the assimilation system against independent observational datasets. Two different experiments, one without data assimilation (\textit{noDA}) and one with data assimilation using a 32-member Ensemble of Data Assimilations (EDA) with 4DEnVar (\textit{32mem}), were validated and evaluated against ground-based and aircraft observations. The observation operators for the independent observations were applied during the data assimilation experiment in monitoring mode (i.e., not assimilated) to the background fields in the \textit{noDA} experiment and to the analysis fields in the \textit{32mem} experiment. The following section explores the effectiveness of the assimilation using these additional validation datasets.

\subsubsection{Pandora}

\begin{figure}
    \centering
    \includegraphics[width=0.5\linewidth]{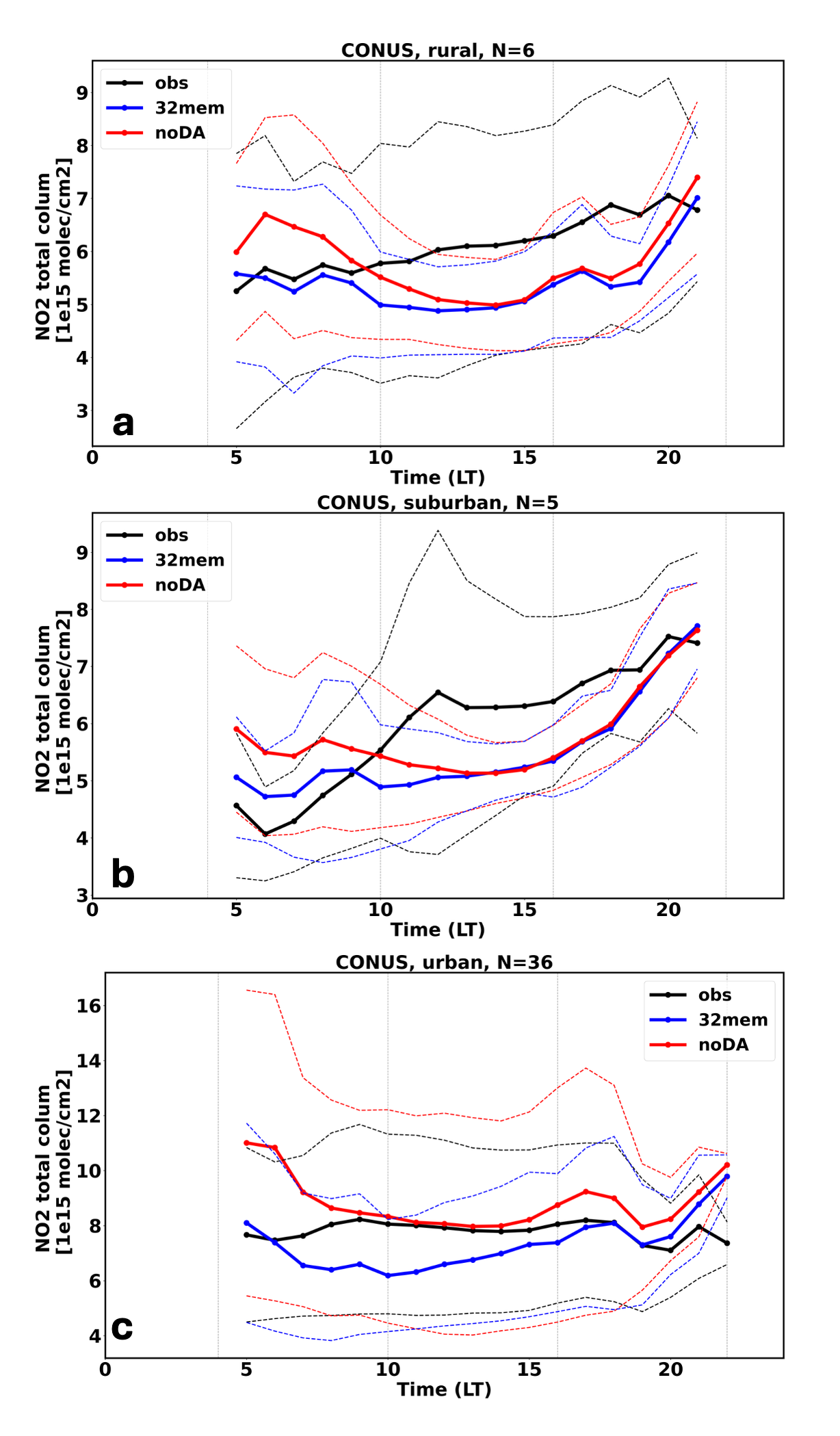}
    \caption{Mean diurnal values of total \ce{NO2} column measured by Pandora over CONUS in (a) rural with 6 stations, (b) suburban with 5 stations, and (c) urban classification with 36 stations. Observation values are in black, and the model-equivalent values of the 32mem and noDA}
    \label{fig:pandora_obs_model}
\end{figure}

Given that $\mathrm{NO}_2$ concentrations and their diurnal variations differ substantially between rural and urban environments, we stratified the analysis by air quality classification (i.e., rural, suburban, and urban) as discussed in section~\ref{subsubsec:pandora}. Figure~\ref{fig:pandora_obs_model} presents the diurnal cycle of total $\mathrm{NO}_2$ column for (a) rural, (b) suburban, and (c) urban Pandora sites. %The spatial distribution and classification of Pandora sites across CONUS are shown in Figure~SM~XX.

Consistent with findings in section~\ref{subsec:obs_space_ana}, the assimilation of TEMPO $\mathrm{NO}_2$ data has a stronger influence during the morning and early afternoon hours. Notably, the assimilation experiment (blue line) demonstrates improved agreement with Pandora observations at rural and suburban sites, while performance over urban sites is degraded during the remainder of the day. This behavior is expected, given the current resolution of the analysis and forecasting system, 1$^\circ$ and 0.25$^\circ$, respectively, which limits the ability to capture fine-scale variability typical of urban environments.

Overall, the assimilation consistently reduces $\mathrm{NO}_2$ column values, indicating a likely overestimation of near-surface $\mathrm{NO}_2$ in the baseline forecast model. The possible underlying causes of this systematic correction will be examined in detail in the following sections.

\subsubsection{Aircraft}

In this section, we evaluate the impact of TEMPO $\mathrm{NO}_2$ assimilation using the aircraft data described in Section~\ref{subsubsec:aircraft}. We performed a comparative analysis of the assimilation and control (noDA) experiments using observations from both remote sensing and \textit{in situ} instruments aboard aircraft platforms.

\begin{figure}
    \centering
    \includegraphics[width=0.5\linewidth]{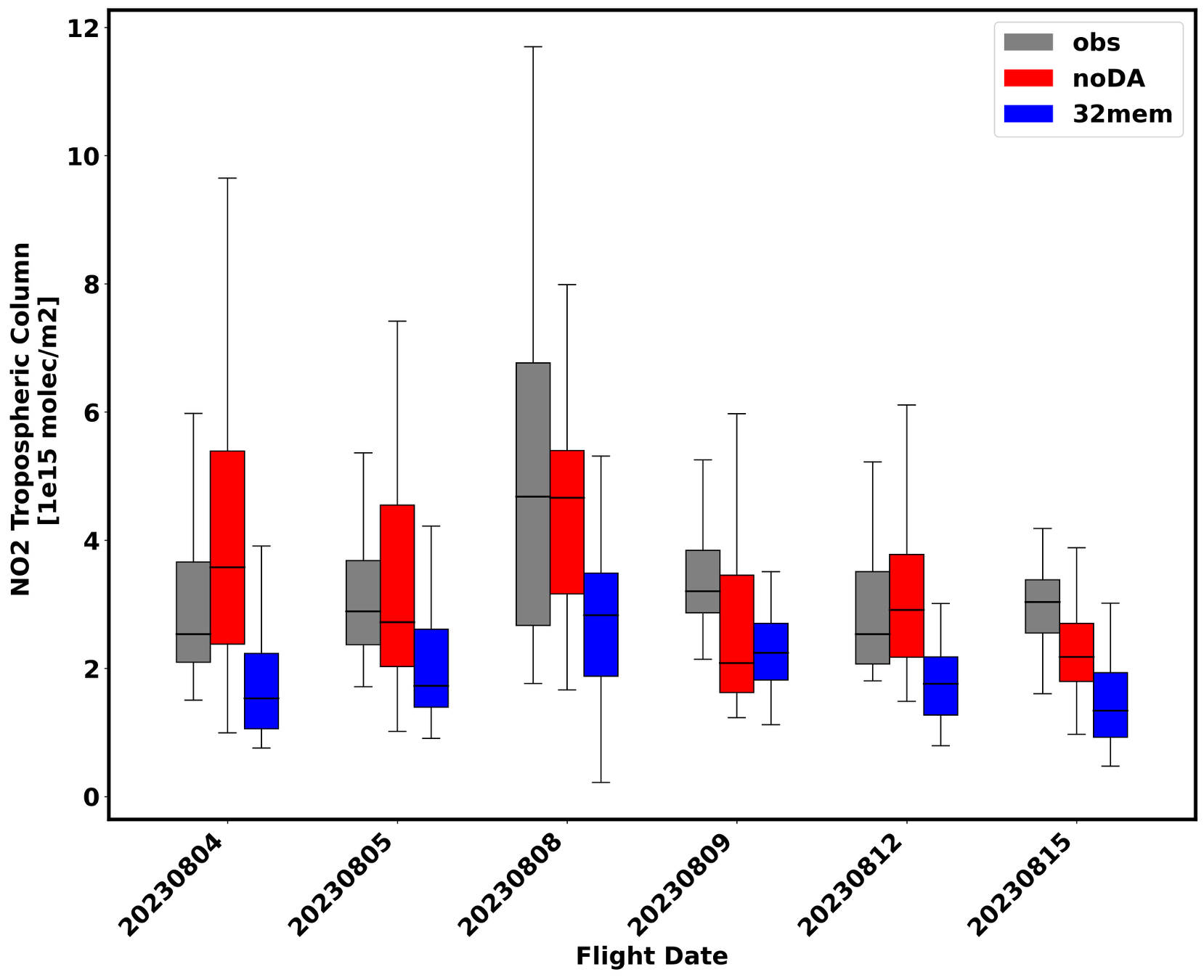}
    \caption{Boxplots of tropospheric \ce{NO2} column measured by GCAS during the STAQS field campaign. Colors grey, red, and blue represent observation, noDA experiment, and 32mem experiment, respectively.}
    \label{fig:gcas_obs_model}
\end{figure}

Figure~\ref{fig:gcas_obs_model} presents the evaluation of the two experiments (i.e., noDA and 32mem) against GCAS measurements (5-min averages), which provide $\mathrm{NO}_2$ column densities below the aircraft flight level (referred to as the tropospheric partial column). The assimilation experiment exhibits a general degradation in agreement with GCAS observations, primarily due to a systematic reduction in $\mathrm{NO}_2$ partial column values. This suggests that the assimilation tends to suppress $\mathrm{NO}_2$ concentrations in the lower troposphere, potentially in response to an overestimated prior.

\begin{figure}
    \centering
    \includegraphics[width=0.75\linewidth]{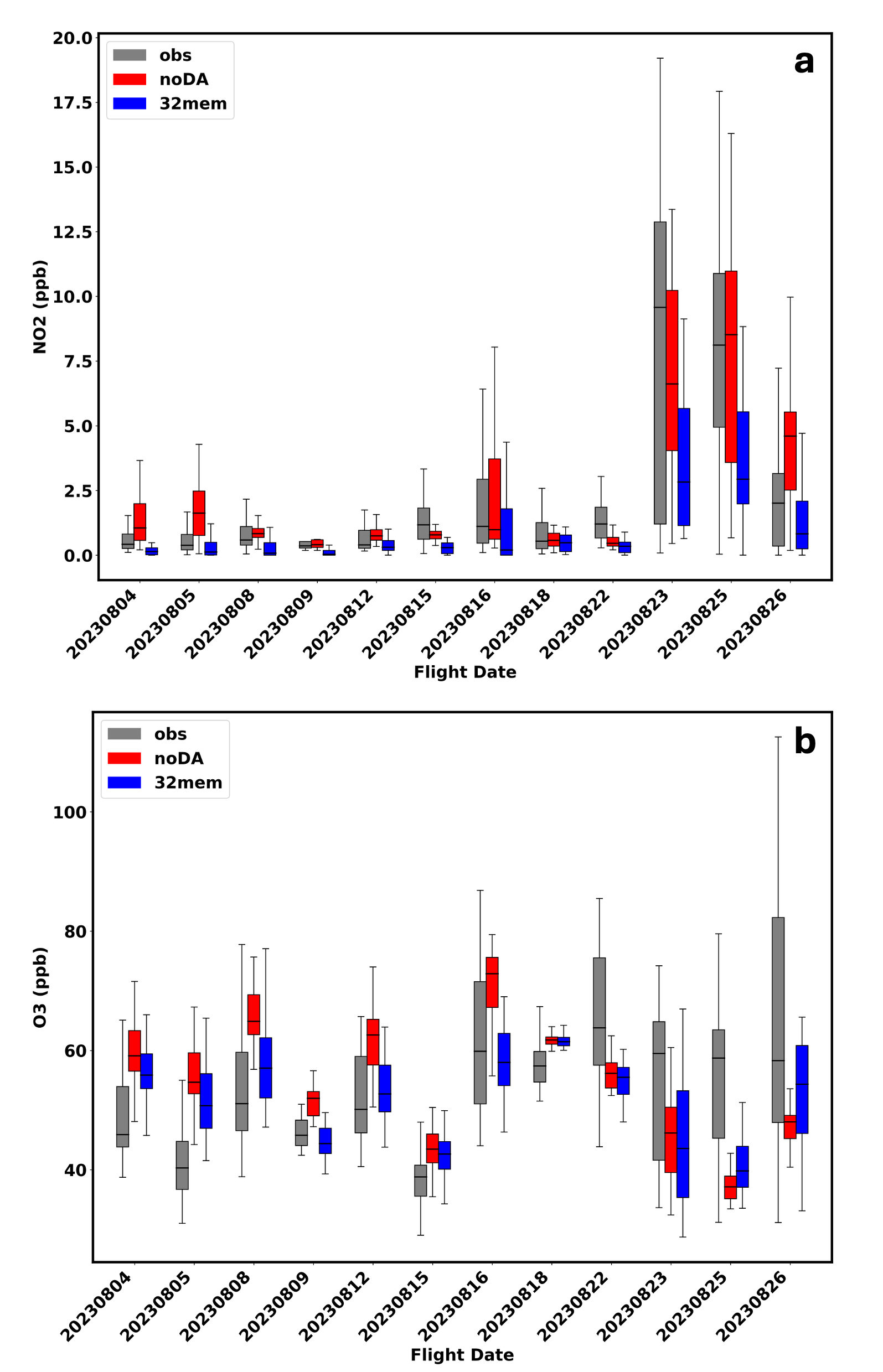}
    \caption{Boxplots of \textit{in situ} \ce{NO2} (a) and ozone (b) concentrations measured during the AEROMMA field campaign. Each box's bottom and top edges represent the 25th (Q1) and 75th (Q3) percentiles, respectively, while the line inside the box denotes the median value. Whiskers extend to ±1 standard deviation from the mean. Colors grey, red, and blue represent observation, noDA experiment, and 32mem experiment, respectively.}
    \label{fig:aeromma_obs_model}
\end{figure}

Figure~\ref{fig:aeromma_obs_model} further illustrates the comparison with AEROMMA \textit{in situ} measurements of $\mathrm{NO}_2$ and ozone. In contrast to the GCAS-based evaluation, the assimilation experiment shows improved performance relative to \textit{in situ} $\mathrm{NO}_2$ measurements. For the majority of the field campaign days, assimilated $\mathrm{NO}_2$ concentrations are reduced compared to the control run, resulting in a better match with observed profiles. A similar pattern is observed for ozone, with consistent improvements in agreement with the \textit{in situ} observations on most days.

These results highlight the complexity of assimilating satellite $\mathrm{NO}_2$ data in a regional model, particularly when evaluated against different types of observational datasets. Discrepancies between the assimilation experiments and aircraft observations may also be influenced by representativeness error, as the model’s resolution of C360 ($\sim$25~km) and the coarser resolution of C90 ($\sim$100~km) used during minimization are insufficient to fully capture the fine-scale variability observed by the aircraft instrument. The broader implications and potential mechanisms driving these impacts will be examined in more detail in Section~\ref{subsec:impact_on_ozone}.

\subsubsection{AirNow}

\begin{figure}
    \centering
    \includegraphics[width=1\linewidth]{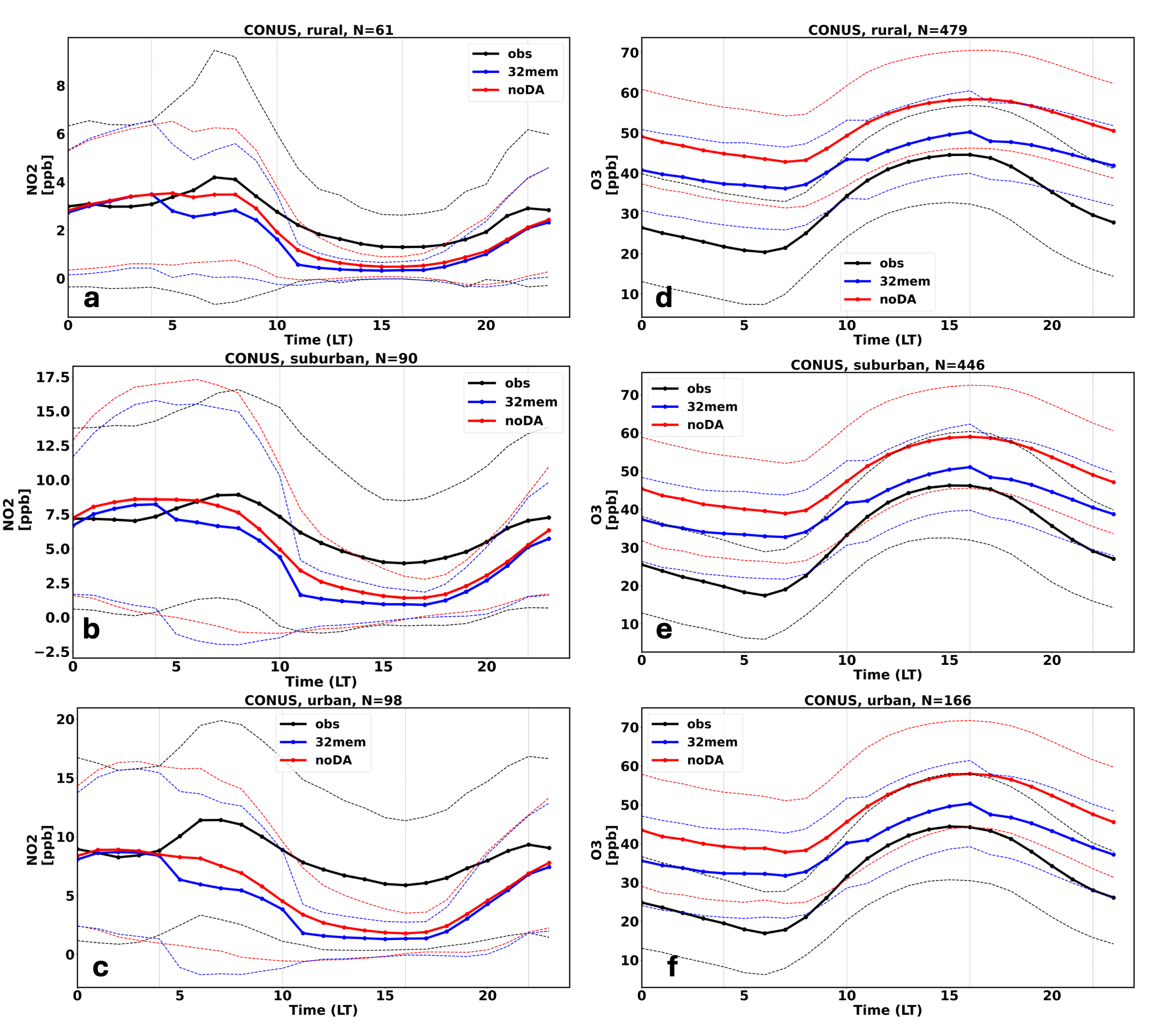}
    \caption{Mean diurnal values of surface \ce{NO2} and ozone concentration averaged over CONUS for rural sites(a and d), suburban sites (b and e), and urban sites (c and f). Observation values are in black, and the model-equivalent values of noDA and 32mem experiments are in red and blue, respectively. Dashed lines indicate ±1 standard deviation range from the mean.  N indicates the number of sites for each plot. 
}
    \label{fig:airnow_diel_obs_model}
\end{figure}

As described in Section~\ref{subsubsec:airnow}, the AirNow network provides measurements of surface $\mathrm{NO}_2$ and ozone mixing ratios across CONUS. We selected the AirNow stations based on their official air quality classifications (rural, suburban, and urban) as provided in the raw observation file metadata. Figure~\ref{fig:airnow_diel_obs_model} presents the diel plots of observed and predicted surface $\mathrm{NO}_2$ and ozone mixing ratios for both noDA and 32mem experiments.

The impact of assimilation on surface $\mathrm{NO}_2$ is generally relatively smaller than its effect on tropospheric or total column values. This is expected given that the assimilation experiment uses tropospheric column observations and not surface measurements. The minimal difference between the noDA and 32mem experiments during nighttime hours is consistent with the lack of assimilated observations at night, and given the short lifetime of $\mathrm{NO}_2$, a quick relaxation to the no assimilation state. Additionally, both experiments overestimate nighttime $\mathrm{NO}_2$ concentrations with varying magnitudes between different study regions (not shown).

During the daytime, the model consistently underestimates surface $\mathrm{NO}_2$ across all site types. As observed with Pandora classified sites, the initial error (observation compared to noDA) is much larger due to the resolution of our forecasting system. Overall, the assimilation of tropospheric $\mathrm{NO}_2$ lowers surface $\mathrm{NO}_2$ mixing ratios. The assimilation further degrades the analysis performance at the surface during the day. This degraded surface $\mathrm{NO}_2$ performance could find its roots in several factors that we will address in the last section of this paper. Despite the degraded $\mathrm{NO}_2$, the assimilation of TEMPO led to significant bias improvement across all \ce{O3} AirNow sites. The impact on \ce{O3} is discussed in the next section.

\subsection{Impacts on ozone}
\label{subsec:impact_on_ozone}

In this study, we did not choose to use the ensemble information to infer ozone from $\mathrm{NO}_2$ observations. This results in strictly zero increments or analysis changes compared to the background for the ozone fields. The ozone changes that we describe in this study are purely the effect of the model chemistry reacting to $\mathrm{NO}_2$ changes in the cycled analysis. Elevated $\mathrm{NO}_2$ acts as a key precursor to ozone formation at the surface and throughout the troposphere, particularly over the CONUS during summer months like August, when photochemical activity is strongest. In the presence of sunlight, $\mathrm{NO}_2$ undergoes photolysis to produce nitric oxide (NO) and atomic oxygen (O), which rapidly react with molecular oxygen (\ce{O2}) to form ozone. This process is sustained through reactions with volatile organic compounds (VOCs), which regenerate $\mathrm{NO}_2$ and maintain the ozone production cycle. Ozone formation is especially active in the morning, when $\mathrm{NO}_2$ levels from early-day anthropogenic emissions are high and solar radiation initiates photolysis. Elevated temperatures and strong solar insolation enhance reaction rates and photolysis efficiency, amplifying ozone production in urban and suburban areas. In chemical transport model forecasts such as GEOS-CF, reducing $\mathrm{NO}_2$ in the initial conditions is expected to weaken or even suppress this cycle, leading to lower ozone concentrations, particularly during the morning hours, by limiting the photochemical production of ozone, especially in VOC-limited environments common in U.S. cities.

\begin{figure}
    \centering
    \includegraphics[width=1\linewidth]{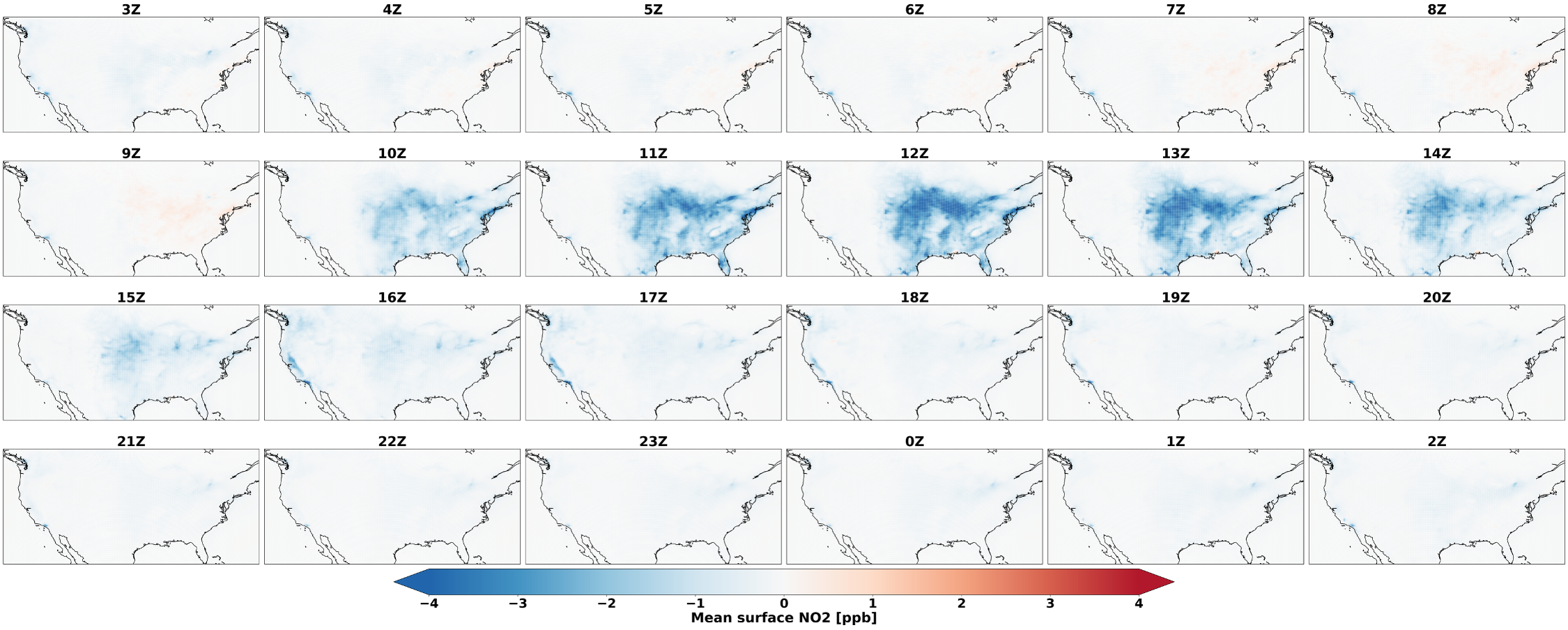}
    \caption{Hourly mean surface \ce{NO2} mixing ratio difference between 32mem and noDA experiments (32mem - noDA) from 4 to 31 August 2023}
    \label{fig:diff_no2_surface}
\end{figure}

In Figure~\ref{fig:diff_no2_surface}, surface $\mathrm{NO}_2$ concentrations show a notable early morning reduction in the 32mem experiment compared to the noDA experiment. These differences between the 32mem and noDA experiments are most pronounced during the early morning assimilation window (09Z to 15Z). This assimilation window coincides with the availability of the early-in-the-day TEMPO observations over the eastern part of the CONUS domain and therefore shows differences in these areas. 

\begin{figure}
    \centering
    \includegraphics[width=1\linewidth]{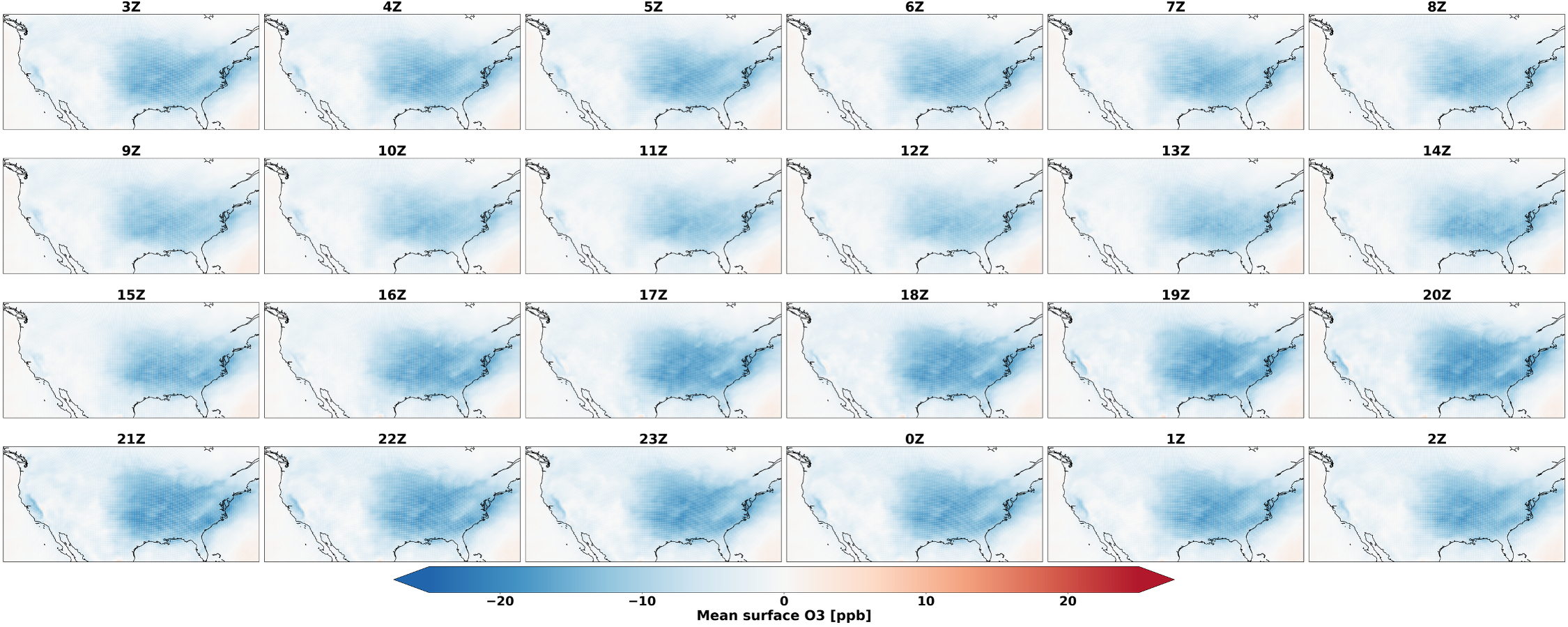}
    \caption{Hourly mean surface \ce{O3} mixing ratio difference between 32mem and noDA experiments (32mem - noDA) from 4 to 31 August 2023}
    \label{fig:diff_o3_surface}
\end{figure}

Figure~\ref{fig:diff_o3_surface} shows the same surface changes but for ozone. Such changes are spreading across all hours of the day and over broader areas compared to $\mathrm{NO}_2$ changes. Given the relatively long lifetime of ozone in the troposphere, ranging from days to weeks, a reduction in surface $\mathrm{NO}_2$ results in propagating lower ozone over time and over regional to continental scales, ultimately leading to a net decrease in ozone across much of the U.S.

This effect indirectly enhances the performance of ozone analyses with respect to independent observations. Given the known positive summertime ozone bias in GEOS-CF over the U.S.~\citep{keller_description_2021}, reducing $\mathrm{NO}_2$ in the initial conditions leads to improved ozone forecasts. As shown in Figure~\ref{fig:airnow_diel_obs_model}, the overall ozone bias is significantly reduced relative to AirNow surface observations. A similar improvement is evident when compared to AEROMMA aircraft measurements (Figure~\ref{fig:aeromma_obs_model}), where a reduction in modeled ozone generally results in better agreement throughout the diurnal cycle. However, this is not always true; on some days, increased ozone appears to yield a better fit, underscoring the need for more detailed diagnostics. Importantly, the observed improvement in ozone, for the "wrong" reasons, namely degraded $\mathrm{NO}_2$ fields, suggests a need for better calibration of VOC emissions and photochemical processes in the model. This represents a key area for future research and highlights the critical role of data assimilation in refining and improving air quality and atmospheric composition forecasting systems.

\section{Discussion and conclusion}
\label{sec:discussion}

In this study, we present the first implementation of TEMPO $\mathrm{NO}_2$ data assimilation within the JEDI framework. TEMPO’s unprecedented geostationary capability delivers hourly tropospheric $\mathrm{NO}_2$ measurements in North America, capturing diurnal variability in emissions and atmospheric chemistry that has historically been inaccessible to regional and global atmospheric composition prediction systems. To fully exploit this high temporal resolution, we demonstrate the use of a four-dimensional ensemble variational (4DEnVar) assimilation system that is explicitly designed to account for time-evolving information within the assimilation window. This approach ensures that the temporal structure of the observations is preserved and dynamically integrated, enabling improved characterization of $\mathrm{NO}_2$ variability at the sub-daily scale.

This work demonstrates, for the first time, the feasibility of applying state-of-the-art data assimilation techniques developed for numerical weather prediction, specifically four-dimensional ensemble variational assimilation (4DEnVar) and the Ensemble of Data Assimilations (EDA), to atmospheric composition and air quality applications. The combined use of 4DEnVar and EDA within the JEDI framework represents a novel configuration in the data assimilation community, enabling the simultaneous representation of flow-dependent background error covariances and time-evolving observational constraints. Our results show that these methods, originally designed for meteorological forecasting, are transferable and effective for assimilating high-frequency chemical observations, providing a robust foundation for next-generation air quality analysis and forecasting systems.

While the implementation of an EDA-based 4DEnVar system for atmospheric composition provides significant scientific advantages, most notably in capturing flow-dependent, time-evolving background errors, it also imposes considerable computational demands. The computational cost of the forecast model dominates the overall cost of the cycling system, and GEOS-CF consumes nearly an order of magnitude more resources than the assimilation components (details not shown). This cost imbalance poses a substantial barrier to scaling such systems for real-time or operational use, especially when ensemble sizes or spatial resolutions are increased. However, recent advances in machine learning, particularly deep learning–based model emulation, offer a transformative opportunity. Ongoing efforts are focused on developing neural network surrogates that replicate the behavior of full-physics models like GEOS-CF with orders-of-magnitude speedups. These emulations aim to preserve critical physical and chemical fidelity while drastically lowering the computational footprint. When integrated into the JEDI framework, such AI-based emulators could make complex assimilation configurations, such as EDA-4DEnVar with high-frequency satellite data, computationally feasible at scale. This approach is under active development and will be demonstrated in a forthcoming study, paving the way for the next generation of efficient, high-resolution atmospheric composition assimilation systems.

This study highlights the versatility and maturity of the JEDI framework as a comprehensive data assimilation system for atmospheric composition. Beyond enabling state-of-the-art analysis and forecast production, JEDI also supports direct evaluation against independent observational datasets, including ground-based, airborne, and satellite platforms, through its unified observation operator infrastructure. This capability eliminates the need for third-party tools or post-processing scripts, streamlining validation workflows, and ensuring methodological consistency across assimilation and evaluation stages. The seamless integration of assimilation and diagnostics within a single framework underscores the potential of JEDI as a robust, end-to-end system for monitoring and forecasting atmospheric composition.

We conducted a comprehensive evaluation of the performance of the assimilation system using both assimilated observations and a diverse suite of independent measurements. This dual-layered assessment allowed us to quantify the direct impact of assimilation and to explore its consistency across observational platforms. Although improvements were observed compared to high-quality reference data sets such as Pandora and AEROMMA measurements, other independent datasets, most notably STAQS and AirNow, revealed systematic degradation in the analysis fields. These contrasting outcomes underscore the complexity of evaluating chemical data assimilation systems, particularly in the presence of differing spatial scales, temporal sampling, and observational uncertainties. The results highlight the importance of a multiperspective validation approach and demonstrate the need for continued refinement of both observation biases and data assimilation components to achieve balanced performance across heterogeneous datasets.

A notable feature of the assimilation results is the pronounced morning reduction in $\mathrm{NO}_2$ concentrations across the eastern CONUS, extending from the East Coast to the Rockies. This signal is directly linked to the assimilation of early-day TEMPO observations, which appear to systematically lower the analyses $\mathrm{NO}_2$ fields during the 09–15Z assimilation window. While this reduction leads to improved agreement with Pandora column measurements, it results in degraded performance when compared to surface-level $\mathrm{NO}_2$ from the AirNow network, particularly at urban sites. These contrasting outcomes highlight the challenges of reconciling column and surface measurements in data assimilation and point to underlying discrepancies in the spatial and vertical representativeness of the observational datasets. Several key factors may contribute to these differences, motivating targeted follow-up studies to further diagnose and improve the assimilation system. They appear as follows:

One possible explanation for the observed inconsistencies is the presence of systematic morning biases between the different observational datasets, particularly between TEMPO, Pandora, and AirNow. While a full characterization of these biases is beyond the scope of this study, the assimilation framework brings them into sharper focus by forcing direct comparisons between datasets with differing vertical sensitivities and temporal sampling. The assimilation-induced morning $\mathrm{NO}_2$ reduction suggests potential high-bias contributions from TEMPO retrievals or low biases in surface-based measurements such as AirNow. These discrepancies underscore the importance of incorporating bias correction strategies into the assimilation system. The JEDI framework supports Variational Bias Correction (VarBC), which could be applied in future studies to mitigate such biases in a statistically consistent manner. However, implementing VarBC requires the careful selection of an anchor dataset, a nontrivial decision in the context of atmospheric composition, where no single dataset provides a universally unbiased reference.

The assimilation-induced reduction in $\mathrm{NO}_2$, particularly during the morning hours, has a clear downstream effect on modeled ozone concentrations due to the photochemical coupling between these species. As shown in Section~\ref{subsec:impact_on_ozone}, lowering $\mathrm{NO}_2$ in the initial conditions leads to reduced ozone formation, which in turn improves agreement with independent ozone observations from both AirNow and AEROMMA. While this suggests a beneficial impact on ozone forecasts, it is important to note that the improvement is achieved indirectly, through $\mathrm{NO}_2$ corrections, rather than via direct assimilation. This points to potential inaccuracies in the representation of VOC emissions and their interplay with $\mathrm{NO}_2$ in the chemical mechanism. Given the VOC-limited regimes common in many urban U.S. regions during summer (e.g. \citep{abdi-oskouei_impact_2022}), underestimation of VOC reactivity may be amplifying the ozone sensitivity to $\mathrm{NO}_2$ reductions. These findings motivate future efforts to refine VOC emission inventories and photochemical parameterizations in GEOS-CF, especially in the context of coupled chemical data assimilation.

While the TEMPO $\mathrm{NO}_2$ retrievals are the most likely contributor to the observed morning degradation in surface $\mathrm{NO}_2$ performance, several other factors within the assimilation system merit further investigation. The current vertical localization configuration may allow spurious correlations to introduce increments that degrade near-surface concentrations. More restrictive vertical localization could help limit the vertical spread of observational influence and reduce unintended impacts at the surface. Additionally, while the system updates the posterior ensemble distribution of concentrations, it does not update the posterior distribution of emissions. Given the static nature of the emission perturbations, it is plausible that the ensemble spread applied to emissions may be too strong, leading to excessive corrections near the surface, especially during sensitive morning hours. These considerations highlight the need for a more balanced and dynamically responsive treatment of emissions within the ensemble framework.

Together, these findings strongly motivate the need for a dual constraint on both concentrations and emissions within the assimilation framework. Simultaneously updating these fields would allow the system to better attribute observational corrections to their appropriate sources, whether due to transport, chemistry, or emissions, and reduce the risk of overcorrecting near-surface concentrations. This represents a novel and ongoing development within the JEDI framework, where new capabilities are being introduced to support joint optimization of atmospheric composition and emission sources. The implementation and evaluation of this dual-assimilation approach will be the focus of a subsequent study.

\section{Acknowledgements}

This work was made possible through the collaborative efforts of many individuals and institutions. We acknowledge valuable contributions from the JCSDA/UCAR team, including Shih-Wei Wei, Sarah Lu, Nate Crossette, Ashley Griffin, Clémentine Hardy Gas, François Hébert, Stephen Herbener, Eric Lingerfelt, Evan Parker, Christian Sampson, Steve Vahl, Fabio Diniz, Ben Johnson, Cheng Dang, Thomas Auligne, Yannick Trémolet, and Benjamin Ruston. We also thank the NASA Global Modeling and Assimilation Office (GMAO) at NASA Goddard Space Flight Center for their support and collaboration, particularly Viral Shah, Emma Knowland (now at NASA Headquarters), Ricardo Todling, Ronald Gelaro, and Steven Pawson.

We gratefully acknowledge the TEMPO principal investigators from the Center for Astrophysics\,Harvard \& Smithsonian—Caroline Nowlan, Gonzalo Gonzalez Abad, and Xiong Lu—for providing critical guidance on instrument data usage. For AEROMMA and STAQS, we thank Brian McDonald (NOAA OAR CSL) and Laura Judd (NASA Langley Research Center) for their leadership and data contributions. Pandora-related data and expertise were generously provided by Thomas Hanisco (NASA GSFC) and Luke Valin (EPA). Additional contributions were made by Benjamin Ménétrier (Met Norway), Daniel Holdaway and Cory Martin (NOAA NWS EMC), and Anna Shlyaeva (UCAR CPAESS), whose technical support and scientific input were essential to the success of this study.

\bibliography{4dEnVar_paper}

\end{document}